\documentclass[a4paper,11pt]{article}
\pdfoutput=1 

\usepackage{jheppub}

\usepackage[T1]{fontenc} 

\newcommand{\AdSS}{\text{AdS}_5\times\text{S}^5}
\newcommand{\su}{\mathfrak{su}}
\newcommand{\psu}{\mathfrak{psu}}
\newcommand{\suCE}{\mathfrak{su}(2|2)_{\text{c.e.}}}
\newcommand{\gen}[1]{\mathbf{#1}}
\newcommand{\comm}[2]{\left[#1,#2\right]}
\newcommand{\acomm}[2]{\left\{#1,#2\right\}}
\newcommand{\half}{\frac{1}{2}}
\newcommand{\ket}[1]{\left|#1\right\rangle}
\newcommand{\Ket}[1]{\big|#1\big\rangle}

\newcommand{\ind}[1]{{\scriptscriptstyle{#1}}}

\title{\boldmath Bethe Ansatz for Quantum-Deformed Strings}

\author[a]{Fiona~K.~Seibold,}
\author[b,c,1]{Alessandro Sfondrini%
\note{IBM Einstein Fellow.}}

\affiliation[a]{Blackett Laboratory, Imperial College London,\\
Prince Consort Road, London, SW7 2AZ, UK}
\affiliation[b]{Institute for Advanced Study\\
Einstein Drive,
Princeton, New Jersey, 08540 USA }
\affiliation[c]{Dipartimento di Fisica e Astronomia, Universit\`a degli Studi di Padova,\\
\& Istituto Nazionale di Fisica Nucleare, Sezione di Padova,\\
via Marzolo 8, 35131 Padova, Italy}

\emailAdd{f.seibold21@imperial.ac.uk}
\emailAdd{alessandro.sfondrini@unipd.it}

\abstract{Two distinct $\eta$-deformations of strings on AdS$_5\times$S$^5$ can be defined; both amount to integrable quantum deformations of the string non-linear sigma model, but only one is itself a superstring background. In this paper we compare their conjectured all-loop worldsheet S~matrices and derive the corresponding Bethe equations. We find that, while the S~matrices are apparently different, they lead to the same Bethe equations. Moreover, in either case the eigenvalues of the transfer matrix, which encode the conserved charges of each system, also coincide.
We conclude that the integrable structure underlying the two constructions is essentially the same. 
Finally, we write down the full Bethe-Yang equations describing the asymptotic spectrum of the superstring background.  
}

\begin{document} 
\maketitle
\flushbottom
\newpage
\section{Introduction and conclusions}
\label{sec:intro}
The duality between type IIB superstrings on the $\AdSS$ background and the  $SU(N)$ $\mathcal{N}=4$ supersymmetric  Yang-Mills theory (SYM)~\cite{Maldacena:1997re} is perhaps the best understood instance of the AdS/CFT correspondence. This is due to its large amount of supersymmetry and to other less manifest symmetries that underlie this setup. In particular, in the planar (large-$N$) limit~\cite{tHooft:1973alw}, this correspondence is integrable, see~\cite{Arutyunov:2009ga, Beisert:2010jr} for reviews. On the gauge theory side, this means that we may treat single-trace local operators as states of a suitable integrable spin chain~\cite{Minahan:2002ve}. By determining an integrable S~matrix for the scattering of magnons we can determine the scaling dimension of long operators and encode it in a set of Bethe ansatz equations~\cite{Beisert:2005fw,Beisert:2005tm,Beisert:2006ez}.
On the string theory side, a similar S~matrix arises on the string worldsheet when considering the scattering of string excitations in a suitable light-cone gauge~\cite{Arutyunov:2006ak,Arutyunov:2006yd}. Once again, a set of Bethe-Yang equations can be derived and it encodes the spectrum of the light-cone Hamiltonian, or equivalently the string energy spectrum in $\text{AdS}_5$, for sufficiently long operators.
The  Bethe-Yang equations are valid only for asymptotically long operators, due to wrapping corrections~\cite{Ambjorn:2005wa}. These can be understood by introducing a mirror theory~\cite{Arutyunov:2007tc} and working out its thermodynamic Bethe ansatz~\cite{Arutyunov:2009zu,Gromov:2009tv,Bombardelli:2009ns,Arutyunov:2009ur}, or by writing a set of ``quantum spectral curve'' equations~\cite{Gromov:2013pga}.
This integrability construction gives a firm grip on the planar spectrum of single-trace operators in $\mathcal{N}=4$ SYM. It also appears that it may be possible to study more general observables, such as three- and higher-point functions~\cite{Basso:2015zoa,Eden:2016xvg,Fleury:2016ykk,Coronado:2018cxj} as well as non-planar ones~\cite{Eden:2017ozn,Bargheer:2017nne}.

Such a success story makes it natural to wonder whether integrability applies to more general instances of AdS/CFT which are less (super)symmetric, or in any case less constrained. Some such instances arise by considering AdS/CFT in lower dimensions, see~\cite{Klose:2010ki,Sfondrini:2014via} for reviews, but it is also possible to deform the $\AdSS$ correspondence itself. Generally speaking, it is easiest to study such deformations from the string side of the duality. Here we are asking whether it is possible to deform the $\AdSS$ supergeometry in such a way that the resulting non-linear sigma model (NLSM) remains classically integrable. It was recently understood~\cite{Delduc:2013fga,Delduc:2013qra,Kawaguchi:2014qwa, Hoare:2014pna} that a very general class of integrable deformations may be described as Yang-Baxter deformations~\cite{Klimcik:2002zj,Klimcik:2008eq}.
Describing in detail, or even listing, all known integrable deformations of $\AdSS$ goes beyond our scope here. Instead, we will focus our attention on certain \emph{quantum deformations} of $\AdSS$ called $\eta$-deformations, whereby the~$\psu(2,2|4)$ superisometry algebra is transformed into a quantum group. This obscures the geometric interpretation of most symmetries, save for the Cartan subalgebra of $\psu(2,2|4)$, and makes it also hard to understand the putative dual theory (the conformal algebra, as well as the Lorentz algebra, are $q$-deformed). Still, it is possible to study the resulting NLSM in detail.

The first $\eta$-deformation of $\AdSS$ was constructed in~\cite{Delduc:2013qra,Delduc:2014kha} and further studied in~\cite{Arutyunov:2013ega,Arutyunov:2015qva}. At tree level, this deformation results in a worldsheet S~matrix which matches with the tensor product of two copies of the integrable S~matrix introduced by Beisert and Koroteev~\cite{Beisert:2008tw} entirely on the basis of symmetry arguments (motivated by the study of $q$-deformations of the Hubbard model). Surprisingly, however, the deformed background \emph{does not satisfy the supergravity equations}~\cite{Arutyunov:2015qva}, even though it does posses $\kappa$-symmetry and obeys a weaker set of conditions~\cite{Arutyunov:2015mqj,Wulff:2016tju}. This fact casts some doubt on whether at the quantum level the construction of~\cite{Delduc:2013qra,Delduc:2014kha} is integrable in the first place, and corresponds to the Beisert-Koroteev S~matrix in the second place; certainly we would expect a Weyl anomaly to appear and spoil $\psu(2,2|4)_q$ invariance. Perhaps a parallel may be drawn with flat-space string theory in non-critical dimension~$D$, where one finds that in light-cone gauge it is possible to define an integrable model starting from the Nambu-Goto action, but that such model does not have the classical $\mathfrak{iso}(1,D-1)$ symmetry~\cite{Dubovsky:2012sh}.

Fortunately, it was recently observed that there exists a \emph{different quantum deformation} of $\AdSS$, which \emph{does satify the supergravity equations} and hence \emph{is a bona-fide superstring background}. Therefore, this latter deformation will presumably be the one of most significance for establishing and studying a deformed AdS/CFT correspondence. The difference between the two deformations is due to the fact that the deforming procedure is formulated in terms of a choice of the Serre-Chevalley basis of $\psu(2,2|4)$. Superalgebras in general, and $\psu(2,2|4)$ in particular, admit different choices as one can pick the simple roots to be of odd/even grading. While the choice of the basis is of no consequence for the definition of the superalgebra, it may lead to different deformations. Indeed, in ref.~\cite{Hoare:2018ngg} it was found that only by picking all simple roots to be odd (Fermionic), the deformed background solves the supergravity equations. It is worth emphasising that the NLSM action for this background coincides  with that of ref.~\cite{Delduc:2013qra,Delduc:2014kha} when all Fermions are put to zero, so that the two constructions are very closely related in quite a concrete sense.
Even more recently, in ref.~\cite{Seibold:2020ywq} it was shown that this ``Femionic'' deformation (as we shall now call it) gives rise to a different S~matrix on the worldsheet at tree level (again differing for the processes involving Fermions) which is classically integrable as expected. Like before, this S~matrix factorises in the product of two copies, and it is possible to conjecture their all-loop form based on $\su(2|2)_q$ symmetry. Specifically, it was found~\cite{Seibold:2020ywq} that this ``Fermionic'' S~matrix is related to the Beisert-Koroteev one by means of a non-diagonal Drinfel'd twist, giving rise to an \textit{a priori} different all-loop S~matrix.

The aim of this article is to determine the Bethe-Yang equations of the ``Fermionic'' S~matrix of ref.~\cite{Seibold:2020ywq}. This is important for at least two reasons: firstly, the Bethe-Yang equations are the stepping stone for the construction of the mirror thermodynamic Bethe ansatz and for the development of even more advanced approaches for computing higher-point functions. Secondly, but importantly, the spectrum is an observable in AdS/CFT, but the worldsheet S~matrix is not. To truly understand the physical difference between the Beisert-Koroteev S~matrix and the ``Fermionic'' one, we should therefore see if they lead to different spectra, and how.

The Bethe equations for the Beisert-Koroteev S~matrix have been derived in the original paper~\cite{Beisert:2008tw} by means of the coordinate Bethe ansatz. 
We find it useful to first repeat that derivation of~\cite{Beisert:2008tw} by means of the \emph{algebraic Bethe ansatz}~\cite{Faddeev:1996iy}, as it was also done in~\cite{Arutyunov:2012ai}. This requires introducing a monodromy matrix $\mathcal{T}_A(\lambda)$ and a transfer matrix $\text{STr}_A[\mathcal{T}_A(\lambda)]$ which is its supertrace. The Bethe equations arise by diagonalising the transfer matrix; moreover, its eigenvalue $\Lambda(\lambda)$ can be expanded in $\lambda\in\mathbb{C}$ to get the conserved quantities of the model. The advantage of this construction, with respect to the coordinate Bethe ansatz, is that it also yields the eigenvalue $\Lambda(\lambda)$; its drawback is that it is somewhat more involved. A convenient way to significantly reduce the computational complexity of the problem is to \textit{assume} that the transfer matrix may be diagonalised, and compute its putative eigenvalue~$\Lambda(\lambda)$. The Bethe equations can then be derived by demanding that $\Lambda(\lambda)$ is regular in $\lambda\in\mathbb{C}$. While this procedure does not prove the existence of the eigenvectors, nor produce their explicit form, it is much simpler and for this reason it is often employed. This is what we shall do for the Beisert-Koroteev and for the ``Fermionic'' S~matrices. We shall also repeat the computation of the coordinate Bethe ansatz for the ``Fermionic'' S~matrix, as a further check.

We find that the Bethe equations for the Beisert-Koroteev and for the ``Fermionic'' S~matrices are identical. This can be seen both from the algebraic and coordinate Bethe ansatz. Moreover, from the algebraic Bethe ansatz we also see that eigenvalues of the two transfer matrices are also identical. 
While our discussion is for the scattering of fundamental particles it is clear that the same will hold for bound states, as those can be obtained by the fusion procedure which is independent from the basis used. At most, the two constructions may differ by the explicit form of the bound-state representation when expressed in terms of the constituent-states basis.
All this strongly suggests that the two resulting integrable models are equivalent. This is perhaps not entirely surprising in view of what happens for lower-dimensional AdS/CFT setups.
Namely, for $\text{AdS}_3 \times \text{S}^3 \times \text{T}^4$~\cite{Seibold:2019dvf}, there is only one deformed S~matrix to begin with, due to the minimal rank of the symmetry algebra~\cite{Hoare:2014oua,Seibold:2021lju}.

To obtain the Bethe-Yang equations for the full superstring theory we must consider two copies of the Fermionic S~matrix. As it turns out~\cite{Seibold:2020ywq} the quantum-deformation parameter~$q$ enters the two copies of the S~matrix in different (in fact, opposite) ways. It was argued in ref.~\cite{Seibold:2020ywq} that this is the case both in the Fermionic setup and in the original one~\cite{Arutyunov:2013ega,Arutyunov:2015qva}.
We conclude that these two apparently inequivalent backgrounds---one of which is not even a superstring background---admit the same asymptotic spectrum under the assumption that they are both all-loop integrable and that the $\su(2|2)_q$ symmetry is non-anomalous. In fact the same should then hold also for the full spectrum, so that the mirror thermodynamic Bethe ansatz~\cite{Arutyunov:2012zt,Arutyunov:2012ai,Arutyunov:2014wdg} and quantum spectral curve~\cite{Klabbers:2017vtw} derived for the model based on the Beisert-Koroteev should matrix should also hold for the ``Fermionic'' deformation. 

The equivalence of the two models might be traced back to the rather constraining assumptions that they both are integrable with $\su(2|2)_q$ symmetry at all loops (or more precisely, with $\su(2|2)_q\oplus\su(2|2)_{1/q}$ symmetry for the whole light-cone-gauge-fixed model). Perhaps this is restrictive enough to guarantee the same integrable structure and eventually the same spectrum. In a sense, the more subtle point is if the construction of refs.~\cite{Delduc:2013qra,Delduc:2014kha} results in a quantum integrable model with the Beisert-Koroteev S~matrix; this may be possible in view of what happens for non-critical flat-space strings~\cite{Dubovsky:2012sh}, but it is not a foregone conclusion. If both constructions are all-loop integrable, subtle differences may \textit{a priori} still appear either in their dressing factors or in the identification of the string tension and deformation parameter beyond tree-level.
It would be interesting to understand this point in more detail, which will likely require a detailed study of the quantisation of the classical integrable construction. In any case we are quite confident, based on the well-established AdS/CFT integrability paradigm, that the supergravity background obtained as a ``Fermionic'' deformation~\cite{Hoare:2018ngg} gives rise to an all-loop integrable model, whose asymptotic spectrum is given by the equations we derive in this paper. It is therefore natural to take this model as a starting point for trying to define a $q$-deformed holographic paradigm, an important goal which we intend to pursue in the future.

This article is structured as it follows. In section~\ref{sec:Smatrix} we review some essential facts about the Beisert-Koroteev S~matrix and about the ``Fermionic'' one. In section~\ref{sec:ABA} we derive the algebraic Bethe ansatz first for the Beisert-Koroteev S~matrix and then for the ``Fermionic'' one, and we argue their equivalence; for completeness, we also report the derivation of the coordinate Bethe ansatz, which can be found in appendix~\ref{app:CBA}. Finally, in section~\ref{sec:result} we write down the complete Bethe-Yang equations for the ``Fermionic'' deformation.

\section{Two quantum-deformed S matrices}
\label{sec:Smatrix}
Before any deformation, the superisometries of the $\AdSS$ background are given by the $\psu(2,2|4)$ superalgebra. The worldsheet S~matrix does not manifestly possess all such symmetries, as it is defined after gauge fixing the model, see refs.~\cite{Arutyunov:2009ga} for a review. The residual algebra is given by two copies of $\su(2|2)$ which undergo a further central extension~\cite{Beisert:2005tm,Arutyunov:2006yd}; we will denote the centrally extended algebra as $\suCE^{\oplus2}$ and we refer the reader to \textit{e.g.}~\cite{Arutyunov:2009ga} for its construction. It turns out that the short (\textit{i.e.}, supersymmetric) representations of $\suCE^{\oplus2}$ which describe the fundamental particles (eight Bosons and eight Fermions) of the lightcone-gauge-fixed $\AdSS$ superstring are given by tensor products of short representations of \emph{a single copy} of $\suCE$, and that as a consequence \emph{the full S~matrix factorises} as
\begin{equation}
    \mathcal{S}(p_1,p_2)_{\AdSS} = 
    \Sigma(p_1,p_2)\;
    \mathcal{R}(p_1,p_2)\check{\otimes}\mathcal{R}(p_1,p_2)\,,
\end{equation}
where $\Sigma(p_1,p_2)$ is a scalar \textit{dressing factor}~\cite{Beisert:2006ez,Hoare:2011wr},  and the check denotes an appropriately graded tensor product (see again~\cite{Arutyunov:2009ga}). Therefore, for many purposes including the derivation of the Bethe-Yang equations, it is sufficient to focus on the $\suCE$-invariant R~matrix $\mathcal{R}(p_1,p_2)$. This intertwines two short representations of $\suCE$, which are of the form $(\textbf{2}|\textbf{2})$, and therefore can be represented as a $16\times 16$ matrix. From now on, when talking about the S~matrix, the particles, \textit{etc.}, we will have in mind a single copy of $\suCE$, its $(\textbf{2}|\textbf{2})$ representations, and the matrix $\mathcal{R}(p_1,p_2)$.

\subsection{The Beisert-Koroteev S matrix}
\label{sec:Smatrix:BK}
Let us review the results of ref.~\cite{Beisert:2008tw}. We will do this following the notation of \cite{Beisert:2011wq}, which is  also the convention used in \cite{Seibold:2020ywq}. We however use different ranges for the indices $a$ and $\alpha$.

\subsubsection{Algebra and deformation}
\label{sec:Smatrix:algebra}
We start by recalling that the $\su(2|2)$ Lie superalgebra has even generators $\gen{L}^\alpha{}_\beta$ ($\alpha=1,2$), $\gen{R}^a{}_b$ ($a=3,4$) subject to $\gen{L}^\alpha{}_\alpha=\gen{R}^a{}_a=0$ (for the two $\su(2)$'s) and $\gen{C}$, as well as odd generators $\gen{Q}^\alpha{}_a$ and $\gen{S}^a{}_\alpha$ (corresponding to supercharges and superconformal charges, respectively). The commutation relations are given by
\begin{equation}
\begin{aligned}
&\comm{\gen{R}^a{}_b}{\gen{R}^c{}_d}=
\delta^c_b\gen{R}^a{}_d
-\delta^a_d\gen{R}^c{}_b\,,\qquad
&&
\comm{\gen{L}^\alpha{}_\beta}{\gen{L}^\gamma{}_\delta}=
\delta^\gamma_\beta\gen{L}^\alpha{}_\delta
-\delta^\alpha_\delta\gen{L}^\gamma{}_\beta\,,
\\
&\comm{\gen{R}^a{}_b}{\gen{Q}^\gamma{}_d}=
-\delta^a_d\gen{Q}^\gamma{}_b
+\half \delta^a_b\gen{Q}^\gamma{}_d\,,
&&
\comm{\gen{L}^\alpha{}_\beta}{\gen{Q}^\gamma{}_d}=
\delta^\gamma_\beta\gen{Q}^\alpha{}_d
-\half \delta^\alpha_\beta\gen{Q}^\gamma{}_d\,,
\\
&\comm{\gen{R}^a{}_b}{\gen{S}^c{}_\delta}=
\delta^c_b\gen{S}^a{}_\delta
-\half \delta^a_b\gen{S}^c{}_\delta\,,
&&
\comm{\gen{L}^\alpha{}_\beta}{\gen{S}^c{}_\delta}=
-\delta^\alpha_\delta\gen{S}^c{}_\beta
+\half \delta^\alpha_\beta\gen{S}^c{}_\delta\,,
\end{aligned}
\end{equation}
as well as%
\begin{equation}
\acomm{\gen{Q}^\alpha{}_b}{\gen{S}^c{}_\delta}=
\delta^c_b\,\gen{L}^\alpha{}_\delta +\delta^\alpha_\delta\,\gen{R}^c{}_b
+\delta^c_b\delta^\alpha_\delta\,\gen{C}\,.
\end{equation}
This algebra may be defined in terms of the Cartan matrix
\begin{equation}
\label{eq:Cartanmatrix}
A=
\begin{pmatrix}
2 & -1 & 0   \\
-1 & 0 & 1   \\
0 & 1 & -2 
\end{pmatrix}\,.
\end{equation}
More specifically, this choice of Cartan matrix (which we make following~\cite{Beisert:2008tw}) corresponds to choosing the so-called \textit{distinguished} Dynkin diagram for the algebra, where the first and last simple roots are Bosonic and correspond to the raising operators of $\su(2)\oplus\su(2)\subset\su(2|2)$, namely $\gen{e}_1=\gen{R}^4{}_3$ and $\gen{e}_3=\gen{L}^1{}_2$, while the remaining simple root is odd, namely $\gen{e}_2=\gen{Q}^2{}_4$, so that the Serre-Chevalley basis is
\begin{equation}
\begin{aligned}
&\gen{h}_1=\gen{R}^4{}_4-\gen{R}^3{}_3  \,,&&
\gen{e}_1=\gen{R}^4{}_3\,,&&
\gen{f}_1=\gen{R}^3{}_4\,,
\\
&\gen{h}_2=-\gen{C}-\half\gen{h}_1-\half\gen{h}_3\,,\qquad&&
\gen{e}_2=\gen{Q}^2{}_4\,,\qquad&&
\gen{f}_2=\gen{S}^4{}_2\,,
\\
&\gen{h}_3=\gen{L}^2{}_2-\gen{L}^1{}_1\,,&&
\gen{e}_3=\gen{L}^1{}_2\,,&&
\gen{f}_3=\gen{L}^2{}_1\,.
\end{aligned}
\end{equation}
Finally, we can get $\suCE$ by considering the twofold central extension
\begin{equation}
\acomm{\gen{Q}^{\alpha}{}_{b}}{\gen{Q}^{\gamma}{}_{d}}
=\varepsilon^{\alpha\gamma}\varepsilon_{bd}\,\gen{P}\,, \qquad
\acomm{\gen{S}^{a}{}_{\beta}}{\gen{S}^{c}{}_{\delta}}
=\varepsilon^{ac}\varepsilon_{\beta\delta}\,\gen{K}\,.
\end{equation}

Starting from the above algebra, or more precisely from the universal enveloping algebra of its complexification, we may define the quantum group $\su(2|2)_q$ in terms of the parameter $q\in\mathbb{C}$. In practice we will be interested in $q \in \mathbb{R}$. The deformed $q$-commutators are defined in terms of the entries of~$A$~\eqref{eq:Cartanmatrix} as it follows~\footnote{In this equation use the bracket notation for both commutator and anti-commutator. In particular, $\comm{\gen{e}_2}{\gen{f}_2} = \gen{e}_2 \gen{f}_2 + \gen{f}_2 \gen{e}_2$.}
\begin{equation}
 q^{\gen{h}_j} \gen{e}_k = q^{+A_{jk}} \gen{e}_k q^{\gen{h}_j}\,, \quad
  q^{\gen{h}_j} \gen{f}_k = q^{-A_{jk}} \gen{f}_k q^{\gen{h}_j}\,,\quad
  \comm{\gen{e}_j}{\gen{f}_k} = d_j \delta_{jk} \frac{q^{\gen{h}_j}-q^{-\gen{h}_j}}{q-q^{-1}}\,,
\end{equation}
where the symmetrisers are
\begin{equation}
d_1 = +1\,, \qquad d_2=d_3=-1\,,
\end{equation}
and for the Cartan elements we have simply
\begin{equation}
    q^{\gen{h}_j} q^{\gen{h}_k} = q^{\gen{h}_k} q^{\gen{h}_j}\,.
\end{equation}
The Serre relations may be found in ref.~\cite{Beisert:2008tw}.
There are three central elements
\begin{equation}
\label{eq:central_oxo}
\begin{aligned}
\gen{C} &= -\gen{h}_2 - \frac{1}{2} ( \gen{h}_1 + \gen{h}_3)\,, \\
\gen{P}&= \gen{e}_1 \gen{e}_2 \gen{e}_3 \gen{e}_2 + \gen{e}_2 \gen{e}_3 \gen{e}_2 \gen{e}_1 + \gen{e}_3 \gen{e}_2 \gen{e}_1 \gen{e}_2 + \gen{e}_2 \gen{e}_1 \gen{e}_2 \gen{e}_3 - (q+q^{-1}) \gen{e}_2 \gen{e}_1 \gen{e}_3 \gen{e}_2\,, \\
\gen{K}&= \gen{f}_1 \gen{f}_2 \gen{f}_3 \gen{f}_2 + \gen{f}_2 \gen{f}_3 \gen{f}_2 \gen{f}_1 + \gen{f}_3 \gen{f}_2 \gen{f}_1 \gen{f}_2 + \gen{f}_2 \gen{f}_1 \gen{f}_2 \gen{f}_3 - (q+q^{-1}) \gen{f}_2 \gen{f}_1 \gen{f}_3 \gen{f}_2\,.
\end{aligned} 
\end{equation}
The centrally extended algebra is promoted to a Hopf algebra with coproduct
\begin{equation}
\label{eq:coproduct1_oxo}
\begin{aligned}
\Delta(\gen{h}_j) &= \gen{h}_j \otimes \gen{1} + \gen{1} \otimes \gen{h}_j~, \\
\Delta(\gen{e}_j) &= \left\{
\begin{aligned}
&\gen{e}_{j} \otimes \gen{1}  + q^{-\gen{h}_{j}} \otimes \gen{e}_{j} &\qquad &j=1,3 ~, \\
&\gen{e}_j \otimes \gen{U}^{-1/2}  + q^{-\gen{h}_j} \gen{U}^{1/2} \otimes \gen{e}_j&\qquad &j=2\,,
\end{aligned}
\right. \\
\Delta(\gen{f}_j) &= \left\{
\begin{aligned}
&\gen{f}_j \otimes q^{\gen{h}_j}  + \gen{1}  \otimes \gen{f}_j &\qquad &\,\,\,j=1,3\,, \\
&\gen{f}_j \otimes q^{\gen{h}_j} \gen{U}^{1/2}  + \gen{U}^{-1/2}  \otimes \gen{f}_j &\qquad &\,\,\,j=2\,,
\end{aligned}
\right.
\end{aligned}
\end{equation}
where $\gen{U}$ is a central element whose eigenvalue we shall specify later. We used the convention of~\cite{Seibold:2020ywq}, so that $\gen{U}$ appears symmetrically in the coproduct.
The coproduct for the central elements follows from \eqref{eq:central_oxo} and it reads
\begin{equation}
\begin{aligned}
\Delta(\gen{C}) &= \gen{C} \otimes \gen{1} + \gen{1} \otimes \gen{C}\,, \\
\Delta(\gen{P}) &= \gen{P} \otimes \gen{U}^{-1} + q^{2 \gen{C}} \gen{U}\otimes \gen{P}\,, \\
\Delta(\gen{K}) &= \gen{K} \otimes \gen{U}\,q^{-2 \gen{C}} + \gen{U}^{-1} \otimes \gen{K}\,,\\
\Delta(\gen{U}) &= \gen{U} \otimes \gen{U}\,.
\end{aligned}
\end{equation}
The opposite coproduct is defined in term of the graded permutation operator $\Pi^g$,%
\footnote{The graded permutation operator of two elements in a tensor product returns the swapped elements, times an additional minus sign if they are both Fermions.}
\begin{equation}
\Delta_\text{op}(\gen{X}) = \Pi^g \Delta(\gen{X}) \Pi^g\,,
\end{equation}
where $\gen{X}$ is any generator.
The conditions
\begin{equation}
\Delta_\text{op}(\gen{P}) = \Delta(\gen{P})~, \qquad \Delta_\text{op}(\gen{K}) = \Delta(\gen{K})~,
\end{equation}
impose
\begin{equation}
\gen{P} = \beta_1 \, \gen{U}^{-1} \left( 1- q^{2 \gen{C}} \gen{U}^{2} \right)\,,
\qquad
\gen{K}=  \beta_2 \, \gen{U} \left( q^{-2 \gen{C}} - \gen{U}^{-2} \right)~,
\end{equation}
up to two undetermined complex coefficients $\beta_{1}$ and $\beta_2$.
Since in what follows we will be interested in unitary representations, we shall require $\gen{P}^\dagger =q^{2\gen{C}}\, \gen{K}$ and $\gen{C}\geq0$, from which it follows that $\gen{U}^\dagger\gen{U}=\gen{1}$ and $\beta_1^*=\beta_2$ (recall that we restrict to $q \in \mathbb{R}$). Moreover, notice that redefining all positive simple roots by a phase $\gen{e}_j\to e^{i\xi}\gen{e}_j$ (as well as $\gen{f}_j\to e^{-i\xi}\gen{f}_j$) is an automorphism that can be used to ``rotate'' $\gen{P}$ and $\gen{K}$, see eq.~\eqref{eq:central_oxo}. Therefore, without loss of generality we may set
\begin{equation}
    \beta_1=\beta_2=\beta\in\mathbb{R}\,.
\end{equation}

\subsubsection{The S matrix}
\label{sec:Smatrix:explicitBK}
The Beisert-Koroteev S~matrix $\mathcal{R}$ can be fixed up to an overall dressing factor by requiring that
\begin{equation}
\label{eq:Rcommutes}
\Delta_{\text{op}}(\gen{X})\, \mathcal{R}= \mathcal{R}\, \Delta(\gen{X})\,,
\end{equation}
for all generators~$\gen{X}$.
Introducing the basis for the $(\mathbf{2}|\mathbf{2})$ representation of the aforedefined Hopf algebra
\begin{equation}
    (\psi_\alpha|\phi_a)\,,\qquad \alpha=1,2\,,\quad a=3,4\,,
\end{equation}
where $\psi_\alpha$ are Fermions and $\phi_a$ are bosons, we can spell out the S-matrix entries explicitly. Notice that in our conventions $\mathcal{R}$ does not permute the momenta. Namely we have, for instance
\begin{equation}
    \mathcal{R} \ket{\phi_a(p_1) \phi_a(p_2)} = A(p_1,p_2) \ket{\phi_a(p_1) \phi_a(p_2)}\,,
\end{equation}
and so on. In what follows, we will omit the dependence on the momenta $p_1,p_2$ both in the states and in the S-matrix elements. When necessary we will indicate the dependence on $p_1,p_2$ by a subscript $1$ or $2$. We then write
\begingroup
\allowdisplaybreaks
\begin{align}
\label{eq:Smat_dist1}
\mathcal{R} \Ket{\phi_a \phi_a} &= A \Ket{\phi_a \phi_a}\\
\mathcal{R} \Ket{\psi_\alpha \psi_\alpha} &= -D \Ket{\psi_\alpha \psi_\alpha}\,, \\
\mathcal{R} \Ket{\phi_a \psi_\alpha} &= G \Ket{\phi_a \psi_\alpha} + H_{a\alpha} \Ket{\psi_\alpha \phi_a}\,,\\
\mathcal{R} \Ket{\psi_\alpha \phi_a} &= L  \Ket{\psi_\alpha \phi_a} + K_{\alpha a}\Ket{\phi_a \psi_\alpha}\,,\\
\nonumber
\mathcal{R} \Ket{\phi_3 \phi_4} &= \frac{A-B}{q+q^{-1}} \Ket{\phi_3 \phi_4} + \frac{\hat{a}_1}{\hat{a}_2} \frac{q A + q^{-1} B}{q+q^{-1}} \Ket{\phi_4 \phi_3}\\
&\qquad\qquad\qquad\qquad\qquad\qquad+ \frac{ \hat{b}_1}{ \hat{a}_2}  \frac{q C}{q+q^{-1}}\Ket{\psi_1 \psi_2} - \frac{\hat{b}_2}{\hat{a}_2} \frac{C}{q+q^{-1}} \Ket{\psi_2 \psi_1}, \\
\nonumber
\mathcal{R} \Ket{\phi_4 \phi_3} &= \frac{\hat{a}_2}{\hat{a}_1}\frac{q^{-1} A + q B}{q+q^{-1}} \Ket{\phi_3 \phi_4} +  \frac{A-B}{q+q^{-1}}  \Ket{\phi_4 \phi_3}\\
&\qquad\qquad\qquad\qquad\qquad\qquad-  \frac{ \hat{b}_1}{ \hat{a}_1}  \frac{q^2 C}{q+q^{-1}}\Ket{\psi_1 \psi_2} + \frac{\hat{b}_2}{\hat{a}_1} \frac{q C}{q+q^{-1}} \Ket{\psi_2 \psi_1}, \\
\nonumber
\mathcal{R} \Ket{\psi_1 \psi_2} &= -\frac{D-E}{q+q^{-1}} \Ket{\psi_1 \psi_2} - \frac{\hat{b}_2}{\hat{b}_1} \frac{q D + q^{-1} E}{q+q^{-1}} \Ket{\psi_2 \psi_1}\\
&\qquad\qquad\qquad\qquad\qquad\qquad-   \frac{ \hat{a}_2}{ \hat{b}_1}  \frac{q^{-1} F}{q+q^{-1}}\Ket{\phi_3 \phi_4} +\frac{\hat{a}_1}{\hat{b}_1} \frac{q^{-2} F}{q+q^{-1}} \Ket{\phi_4 \phi_3}, \\
\nonumber
\mathcal{R} \Ket{\psi_2 \psi_1} &= -  \frac{\hat{b}_1}{\hat{b}_2} \frac{q^{-1} D + q E}{q+q^{-1}} \Ket{\psi_1 \psi_2}  -\frac{D-E}{q+q^{-1}} \Ket{\psi_2 \psi_1}\\
&\qquad\qquad\qquad\qquad\qquad\qquad+ \frac{ \hat{a}_2}{ \hat{b}_2}  \frac{F}{q+q^{-1}}\Ket{\phi_3 \phi_4} - \frac{\hat{a}_1}{\hat{b}_2} \frac{q^{-1} F}{q+q^{-1}} \Ket{\phi_4 \phi_3},
\end{align}
\endgroup
with
\begin{equation}
\begin{aligned}
H_{31} = \frac{\hat{b}_1}{\hat{b}_2} H\,, \qquad H_{32}=H\,, \qquad H_{41}= \frac{\hat{a}_2 \hat{b}_1}{\hat{a}_1 \hat{b}_2} H\,, \qquad H_{42} = \frac{\hat{a}_2}{\hat{a}_1} H\,, \\
K_{13} = \frac{\hat{b}_2}{\hat{b}_1} K~, \qquad K_{23}=K\,, \qquad K_{14}= \frac{\hat{a}_1 \hat{b}_2}{\hat{a}_2 \hat{b}_1} K\,, \qquad K_{24} = \frac{\hat{a}_1}{\hat{a}_2} K\,,
\end{aligned}
\end{equation}
and
\begingroup
\allowdisplaybreaks
\begin{align}
A &=  \frac{U_1 V_1}{U_2 V_2} \frac{x_2^+ - x_1^-}{x_2^- - x_1^+}\,, \\
B &=  \frac{U_1 V_1}{U_2 V_2} \frac{x_2^+ - x_1^-}{x_2^- - x_1^+} \left( 1-(q+q^{-1})q^{-1} \frac{x_2^+ - x_1^+}{x_2^+ - x_1^-} \frac{x_2^- - 1/x_1^+}{x_2^- - 1/x_1^-}\right)\,,\\
C &= - (q+q^{-1}) \frac{ \gamma_1 \gamma_2 U_1 V_1}{\alpha q^{3/2} U_2^2 V_2^2}\frac{x_1^-}{x_1^+} \frac{x_1^+ - x_2^+}{(x_1^+ - x_2^-)(1- x_1^- x_2^-)}~, \\
D &= -1\,, \\
E &= -  \left( 1-(q+q^{-1}) \frac{1}{q U_2^2 V_2^2} \frac{x_2^+ - x_1^+}{x_2^- - x_1^+} \frac{x_2^+ - 1/x_1^-}{x_2^- - 1/x_1^-}\right)\,,\\
F &= -  (q+q^{-1}) \frac{\alpha U_1^2 V_1^2}{q^{1/2} U_2 V_2 \gamma_1 \gamma_2} \frac{x_1^-}{x_1^+} \frac{(x_1^- - x_1^+)(x_2^+ - x_1^+)(x_2^+ - x_2^-)}{(x_2^- - x_1 ^+)(1-x_1^- x_2^-)}\,, \\
G &=  \frac{1}{q^{1/2} U_2 V_2} \frac{x_2^+ - x_1^+}{x_2^- - x_1^+}\,, \\
H &=   \frac{\gamma_1}{\gamma_2} \frac{x_2^+ - x_2^-}{x_2^- - x_1^+}\,, \\
K &=  \frac{U_1 V_1}{U_2 V_2} \frac{\gamma_2}{\gamma_1} \frac{x_1^+ - x_1^-}{x_2^- - x_1^+}\,, \\
L &=  U_1 V_1 q^{1/2} \frac{x_2^- - x_1^-}{x_2^- - x_1 ^+}\,.
\end{align}
\endgroup
We will comment below on the various parameters that appear in these formulae.

\subsubsection{Parametrisation for physical particles}
\label{sec:Smatrix:parametrisation}
Above we have introduced the short-hand $V_j$ to indicate the eigenvalue of $q^{\gen{C}}$ on the $j$-th particle, \textit{e.g.}~ $q^{\gen{C}}\ket{\phi_a(p_j)}=V_j\ket{\phi_a(p_j)}$. The eigenvalue $V$ is related to the eigenvalue $U$ of  $\gen{U}$ by the closure condition
\begin{equation}
\label{eq:closure}
\xi^2 (U-U^{-1})^2 - (V-V^{-1})^2 + (1-\xi^2) (q^{1/2}-q^{-1/2})^2 =0\,,
\end{equation}
where
\begin{equation}
\xi = -i\frac{\beta (q-q^{-1})}{\sqrt{1-\beta^2 (q-q^{-1})^2}}\,.
\end{equation}
Moreover, $U$ and $V$ are related to the Zhukovsky variables $x^{\pm}$ through
\begin{equation}
\label{eq:UV}
U^2 = q^{-1} \frac{x^+ + \xi}{x^- + \xi} =  q \frac{1/x^-+\xi}{1/x^+ + \xi}~, \qquad
V^2 = q^{-1} \frac{1+ x^+ \xi}{1+ x^- \xi} = q \frac{\xi/x^- + 1}{\xi/x^+ +1}\,,
\end{equation}
and hence in the $x^\pm$ variables the closure condition  becomes
\begin{equation}
\label{eq:eqq}
q^{-1} \left( x^+ + \frac{1}{x^+} \right)- q \left( x^- + \frac{1}{x^-} \right) -(q-q^{-1}) \left(\xi + \frac{1}{\xi}\right)=0\,.
\end{equation}
To have an interpretation of this S matrix as (half) the physical S matrix representing the scattering of excitation on the worldsheet of strings moving in deformed space-time  we make the identification
\begin{equation}
V= q^{\omega/2}~, \qquad U = e^{ip/2}\,,
\end{equation}
with $\omega$ and $p$ the energy and momentum of a given excitation, respectively. The parameter $\xi$ is related to the string tension and the deformation parameter $q$. It is such that $\xi \rightarrow 0$ when $q \rightarrow 1$ (undeformed case), with the following well-defined limit
\begin{equation}
(q-q^{-1})(\xi+\xi^{-1}) \rightarrow \frac{i}{\beta}\,.
\end{equation}
The parameter $\beta$ can then be identified with half the string tension.

To ensure physical unitarity of the S~matrix we take the following reality conditions
\begin{equation}
\label{eq:untwistedab}
\hat{a}= q^{1/2}\,, \qquad \hat{b}=q^{-1/2}\,.
\end{equation}
Notice that this then implies $H_{a\alpha}=H$ and $K_{\alpha a}= K$ for all $a,\alpha$. Note also that this replacement somewhat simplifies the explicit form of the S~matrix. Finally, there is the parameter $\gamma(p)$ that enters in the coefficients $C$, $F$, $H$ and $K$ of the S matrix, and encodes the normalisation of the Fermions with respect to the Bosons. 

The S~matrix obeys the Yang-Baxter equation
\begin{equation}
\label{eq:YBE}
    \mathcal{R}_{12}(p_1,p_2)\,
    \mathcal{R}_{13}(p_1,p_3)\,
    \mathcal{R}_{23}(p_2,p_3)=
    \mathcal{R}_{23}(p_2,p_3)\,
    \mathcal{R}_{13}(p_1,p_3)\,
    \mathcal{R}_{12}(p_1,p_2)\,,
\end{equation}
where $\mathcal{R}_{12}(p_1,p_2)=\mathcal{R}(p_1,p_2)\otimes \gen{1}$, $\mathcal{R}_{23}(p_2,p_3)=\gen{1}\otimes\mathcal{R}(p_2,p_3)$, and $\mathcal{R}_{13}(p_1,p_3)=(\Pi^g\otimes\gen{1})\cdot\gen{1}\otimes\mathcal{R}(p_1,p_3)\cdot (\Pi^g\otimes\gen{1})$.
Notice that the S~matrix satisfies the Yang-Baxter equation without the need of adding any twist because we are working in the ``string frame'' of~\cite{Arutyunov:2006yd}.
Finally, it is worth emphasising that when the representation parameters coefficients coincide for the two particles, the S~matrix reduces to (minus) the graded permutation, 
\begin{equation}
\label{eq:Spp}
\mathcal{R}(p,p) = - \Pi^g\,.
\end{equation} 

\subsection{The Fermionic S~matrix}
\label{sec:Smatrix:Fermionic}
There is another interesting definition of a $q$-deformed S~matrix. This emerges~\cite{Hoare:2018ngg} from the Yang-Baxter deformation of the $\AdSS$ sigma model constructed from a fully Fermionic $\psu(2,2|4)$ Dynkin diagram---one where all the simple roots are odd. Additionally, the symmetries of the S~matrix are given by the quantum group constructed from $\suCE$ with the following choice of Cartan matrix,
\begin{equation}
\tilde{A}=
\begin{pmatrix}
0 & +1 & 0   \\
+1 & 0 & -1   \\
0 & -1 & 0 
\end{pmatrix}\,,
\end{equation}
which again corresponds to picking all simple roots among the odd generators. This is the reason why we refer to this construction and to the resulting S~matrix as ``Fermionic''.

\subsubsection{Non-diagonal twist}
\label{sec:Smatrix:twist}
It is possible to repeat the construction of the quantum group and of the S~matrix $\tilde{\mathcal{R}}(p_1,p_2)$, which again follows from~\eqref{eq:Rcommutes}, as it was done in ref.~\cite{Seibold:2020ywq}. There, it was found that the resulting S~matrix is related to the Beisert-Koroteev one~$\mathcal{R}(p_1,p_2)$ by a \emph{non-diagonal Drinfel'd twist}~$\gen{F}$,
\begin{equation}
    \tilde{\mathcal{R}} = \Pi^g\,\gen{F}^{-1}\,\Pi^g\,\mathcal{R}\,\gen{F}\,,
\end{equation}
where
\begin{equation}
    \gen{F}=\gen{1}\otimes\gen{1} - \big(q-q^{-1}\big)\,\gen{U}^{\tfrac{1}{2}}\gen{f}_2\,\otimes\,\gen{U}^{\tfrac{1}{2}}\gen{e}_2\,,
\end{equation}
which is written in terms of the positive and negative roots introduced above.

\subsubsection{Explicit form of the S~matrix}
\label{sec:Smatrix:explicitFermionic}
It is worth writing explicitly the S~matrix resulting from the twist so that we may highlight the differences with the Beisert-Koroteev one (see also \cite{Seibold:2021lju}). We indicate in red the terms due to the twist.
	\begingroup
\allowdisplaybreaks
\begin{align}
\label{eq:Smat_ferm}
\tilde{\mathcal{R}} \Ket{\phi_a \phi_a} &= A \Ket{\phi_a \phi_a}\\
\tilde{\mathcal{R}} \Ket{\psi_\alpha \psi_\alpha} &= -D \Ket{\psi_\alpha \psi_\alpha}\,, \\
\tilde{\mathcal{R}} \Ket{\phi_a \psi_\alpha} &= G \Ket{\phi_a \psi_\alpha} + H_{a\alpha} \Ket{\psi_\alpha \phi_a}\,,\\
\tilde{\mathcal{R}} \Ket{\psi_\alpha \phi_a} &= L  \Ket{\psi_\alpha \phi_a} + K_{\alpha a}\Ket{\phi_a \psi_\alpha}\,,\\
\nonumber
\tilde{\mathcal{R}} \Ket{\phi_3 \phi_4} &= \frac{A-B}{q+q^{-1}} \Ket{\phi_3 \phi_4} +{\color{red}\varphi_{12}} \frac{\hat{a}_1}{\hat{a}_2} \frac{q A + q^{-1} B}{q+q^{-1}} \Ket{\phi_4 \phi_3}\\
&\qquad\qquad\qquad\qquad\qquad\qquad+{\color{red}\frac{g_2}{f_1}}\frac{ \hat{b}_1}{ \hat{a}_2}  \frac{q C}{q+q^{-1}}\Ket{\psi_1 \psi_2} - \frac{\hat{b}_2}{\hat{a}_2} \frac{{\color{red}q^2}C}{q+q^{-1}} \Ket{\psi_2 \psi_1}, \\
\nonumber
\tilde{\mathcal{R}} \Ket{\phi_4 \phi_3} &= {\color{red}\hat{\varphi}_{12}}\frac{\hat{a}_2}{\hat{a}_1}\frac{q^{-1} A + q B}{q+q^{-1}} \Ket{\phi_3 \phi_4} +  \frac{A-B}{q+q^{-1}}  \Ket{\phi_4 \phi_3}\\
&\qquad\qquad\qquad\qquad\qquad\qquad-  \frac{ \hat{b}_1}{ \hat{a}_1}  \frac{q^2 C}{q+q^{-1}}\Ket{\psi_1 \psi_2} + {\color{red}\frac{g_1}{f_2}}\frac{\hat{b}_2}{\hat{a}_1} \frac{q C}{q+q^{-1}} \Ket{\psi_2 \psi_1}, \\
\nonumber
\tilde{\mathcal{R}} \Ket{\psi_1 \psi_2} &= -\frac{D-E}{q+q^{-1}} \Ket{\psi_1 \psi_2} - {\color{red}\varphi_{21}}\frac{\hat{b}_2}{\hat{b}_1} \frac{q D + q^{-1} E}{q+q^{-1}} \Ket{\psi_2 \psi_1}\\
&\qquad\qquad\qquad\qquad\qquad\qquad- {\color{red}\frac{f_1}{g_2}}\frac{ \hat{a}_2}{ \hat{b}_1}  \frac{q^{-1} F}{q+q^{-1}}\Ket{\phi_3 \phi_4} +\frac{\hat{a}_1}{\hat{b}_1} \frac{q^{-2} F}{q+q^{-1}} \Ket{\phi_4 \phi_3}, \\
\nonumber
\tilde{\mathcal{R}} \Ket{\psi_2 \psi_1} &= - {\color{red}\hat{\varphi}_{21}} \frac{\hat{b}_1}{\hat{b}_2} \frac{q^{-1} D + q E}{q+q^{-1}} \Ket{\psi_1 \psi_2}  -\frac{D-E}{q+q^{-1}} \Ket{\psi_2 \psi_1}\\
&\qquad\qquad\qquad\qquad\qquad\qquad+ \frac{ \hat{a}_2}{ \hat{b}_2}  \frac{{\color{red}q^{-2}}F}{q+q^{-1}}\Ket{\phi_3 \phi_4} - {\color{red}\frac{f_2}{g_1}}\frac{\hat{a}_1}{\hat{b}_2} \frac{q^{-1} F}{q+q^{-1}} \Ket{\phi_4 \phi_3},
\end{align}
\endgroup
where now we have
\begin{equation}
\begin{aligned}
H_{31} &= {\color{red}\frac{f_2}{f_1}} \frac{\hat{b}_1}{\hat{b}_2} H\,, &\qquad H_{32}&= H~, &\qquad H_{41}&= \frac{\hat{a}_2}{\hat{a}_1} \frac{\hat{b}_1}{\hat{b}_2} H\,, &\qquad H_{42}&= {\color{red}\frac{g_1}{g_2}} \frac{\hat{a}_2}{\hat{a}_1} H\,, \\
K_{13} &= {\color{red}\frac{f_1}{f_2}} \frac{\hat{b}_2}{\hat{b}_1} K\,, &\qquad K_{23}&= K~, &\qquad K_{14}&= \frac{\hat{a}_1}{\hat{a}_2} \frac{\hat{b}_2}{\hat{b}_1} K\,, &\qquad K_{24}&= {\color{red}\frac{g_2}{g_1}} \frac{\hat{a}_1}{\hat{a}_2} K\,,
\end{aligned}
\end{equation}
and we have introduced the functions
\begin{equation}
\varphi_{12}=\frac{q^{1/2} f_1 x_1^-+q^{-2 C_1-1/2} g_1 x_2^+}{q^{1/2} f_1 x_1^-+q^{-2 C_2-1/2} g_1 x_2^+}\,, \qquad \hat{\varphi}_{12}=\frac{q^{-1/2} g_2 x_1^- +q^{-2 C_2+1/2} f_2 x_2^+}{q^{-1/2} g_2 x_1^- +q^{-2 C_1+1/2} f_2 x_2^+}\,,
\end{equation}
with
\begin{equation}
f_j = 1+\xi/x_j^-~, \qquad g_j = 1+ \xi x_j^+~.
\end{equation}
In this case we also need to amend the definition of the parameters $\hat{a},\hat{b}$ with respect to ~\eqref{eq:untwistedab} to
\begin{equation}
\hat{a}_j = q^{C_j-1/2}\,, \qquad \hat{b}_j = q^{C_j+1/2}\,,
\end{equation}
where $C_j$ is the eigenvalue of the central charge $\gen{C}\ket{p_j}=C_j\ket{p_j}$. Requiring unitarity will also impose a different condition on $\gamma(p)$, but this can be reabsorbed into a rescaling of the Fermions.
It is apparent that the twisted S~matrix is more involved than the Beisert-Koroteev one. In particular, the coefficients $H_{a\alpha}$ and $K_{\alpha a}$ now depend on the indices $a,\alpha$, which signals that the $\su(2)_q$ symmetries are not manifest in this deformation. Indeed, in the Beisert-Koroteev construction, $\gen{e}_1$ and $\gen{e}_3$ where $\su(2)$ positive roots which made more transparent the deformation of those two subalgebras. 
We will encounter again this difference when studying the Bethe ansatz for the Fermionic S~matrix. 

\subsubsection{Redefinition of the one-particle basis}
\label{sec:Smatrix:Fermionicrescaled}
For the construction the Bethe ansatz it will be convenient to implement a one-particle change of basis which rescales
\begin{equation}
   \Ket{\psi_1(p)} \rightarrow  \frac{\hat{b}(p)}{f(p)} \Ket{\psi_1(p)},\qquad
    \Ket{\phi_4(p)} \rightarrow \frac{\hat{a}(p)}{g(p)} \Ket{\phi_4(p)}\,.
\end{equation}
As a consequence, the ``Fermionic'' S~matrix takes the slightly simpler form
\begingroup
\allowdisplaybreaks
\begin{align}
\label{eq:Smat_ferm1}
\tilde{\mathcal{R}} \Ket{\phi_a \phi_a} &= A \Ket{\phi_a \phi_a}\\
\tilde{\mathcal{R}} \Ket{\psi_\alpha \psi_\alpha} &= -D \Ket{\psi_\alpha \psi_\alpha}\,, \\
\tilde{\mathcal{R}} \Ket{\phi_a \psi_\alpha} &= G \Ket{\phi_a \psi_\alpha} + H_{a\alpha} \Ket{\psi_\alpha \phi_a}\,,\\
\tilde{\mathcal{R}} \Ket{\psi_\alpha \phi_a} &= L  \Ket{\psi_\alpha \phi_a} + K_{\alpha a}\Ket{\phi_a \psi_\alpha}\,,\\
\nonumber
\tilde{\mathcal{R}} \Ket{\phi_3 \phi_4} &= \frac{A-B}{q+q^{-1}} \Ket{\phi_3 \phi_4} +{\color{red}\varphi_{12} \frac{g_1}{g_2}}  \frac{q A + q^{-1} B}{q+q^{-1}} \Ket{\phi_4 \phi_3}\\
&\qquad\qquad\qquad\qquad\qquad\qquad+  \frac{q C}{q+q^{-1}}\Ket{\psi_1 \psi_2} - {\color{red}\frac{f_2}{g_2}} \frac{q^2C}{q+q^{-1}} \Ket{\psi_2 \psi_1}, \\
\nonumber
\tilde{\mathcal{R}} \Ket{\phi_4 \phi_3} &= {\color{red}\hat{\varphi}_{12}\frac{g_2}{g_1}}\frac{q^{-1} A + q B}{q+q^{-1}} \Ket{\phi_3 \phi_4} +  \frac{A-B}{q+q^{-1}}  \Ket{\phi_4 \phi_3}\\
&\qquad\qquad\qquad\qquad\qquad\qquad-  {\color{red}\frac{ f_1}{ g_1}}  \frac{q^2 C}{q+q^{-1}}\Ket{\psi_1 \psi_2} + \frac{q C}{q+q^{-1}} \Ket{\psi_2 \psi_1}, \\
\nonumber
\tilde{\mathcal{R}} \Ket{\psi_1 \psi_2} &= -\frac{D-E}{q+q^{-1}} \Ket{\psi_1 \psi_2} - {\color{red}\varphi_{21} \frac{f_2}{f_1}} \frac{q D + q^{-1} E}{q+q^{-1}} \Ket{\psi_2 \psi_1}\\
&\qquad\qquad\qquad\qquad\qquad\qquad-  \frac{q^{-1} F}{q+q^{-1}}\Ket{\phi_3 \phi_4} +{\color{red}\frac{g_1}{f_1}} \frac{q^{-2} F}{q+q^{-1}} \Ket{\phi_4 \phi_3}, \\
\nonumber
\tilde{\mathcal{R}} \Ket{\psi_2 \psi_1} &= - {\color{red}\hat{\varphi}_{21} \frac{f_1}{f_2}}  \frac{q^{-1} D + q E}{q+q^{-1}} \Ket{\psi_1 \psi_2}  -\frac{D-E}{q+q^{-1}} \Ket{\psi_2 \psi_1}\\
&\qquad\qquad\qquad\qquad\qquad\qquad+  {\color{red}\frac{ g_2}{ f_2}}  \frac{q^{-2}F}{q+q^{-1}}\Ket{\phi_3 \phi_4} -  \frac{q^{-1} F}{q+q^{-1}} \Ket{\phi_4 \phi_3},
\end{align}
\endgroup
where now
\begin{equation}
\begin{aligned}
H_{31} &=  H\,, &\qquad H_{32}&= H~, &\qquad H_{41}&= {\color{red}\frac{f_1 g_2}{f_2 g_1}} H\,, &\qquad H_{42}&=  H\,, \\
K_{13} &=  K\,, &\qquad K_{23}&= K~, &\qquad K_{14}&= {\color{red}\frac{f_2 g_1}{f_1 g_2}} K\,, &\qquad K_{24}&=  K\,.
\end{aligned}
\end{equation}
This is the S matrix that we will later diagonalise using the algebraic and coordinate Bethe ansatz.

\section{Algebraic Bethe ansatz}
\label{sec:ABA}
The Bethe equations for the Beisert-Koroteev S~matrix are known, having been derived along with the S~matrix in ref.~\cite{Beisert:2008tw}. That derivation relied on the coordinate Bethe ansatz, much like the original derivation of the the Bethe equations of the $\suCE$ S~matrix~\cite{Beisert:2005tm}. Here we repeat the derivation of the Beisert-Koroteev Bethe equations in terms of the algebraic Bethe ansatz. For the underformed $\suCE$ S~matrix, the algebraic Bethe ansatz was discussed in ref.~\cite{Martins:2007hb} by mapping the problem to the Hubbard model. For the Beisert-Koroteev S~matrix this was done in~\cite{Arutyunov:2012ai}. We find it useful to repeat this derivation in some detail as a warm-up exercise in view of the diagonalisation of the Fermionic S~matrix of~\cite{Seibold:2020ywq}.
\subsection{Algebraic Bethe ansatz for the Beisert-Koroteev S~matrix}
\label{sec:ABA:BK}
We start by picking a basis for the $(\mathbf{2}|\mathbf{2})$ module which, following~\cite{Martins:2007hb}, we take to be
\begin{equation}
\label{eq:basis}
    \left(\phi_3,\, \psi_1,\, \psi_2,\,\phi_4\right)\,.
\end{equation}
 We introduce an auxiliary module $\mathcal H_A \cong \mathbb{C}^4$ which is spanned by the vectors in~\eqref{eq:basis}. We define the Lax operator $\mathcal{L}_{j,A} (\lambda,p_j)$ in terms of the S~matrix $\mathcal{R}_{A,j}(\lambda,p_j)$ acting on one auxiliary and one physical module:
\begin{equation}
\mathcal{L}_{j,A} (\lambda,p_j) = \mathcal{R}_{A,j}(\lambda,p_j)\,.
\end{equation}
Here we will be using that the S~matrix obeys the Yang-Baxter equations~\eqref{eq:YBE} and that, for identical momenta (\textit{i.e.}\ for $\lambda=p_j$ here) it reduces to (minus) the graded permutation operator, see~\eqref{eq:Spp}.

We choose the vacuum configuration of each site to be given by $\phi_3$,
\begin{equation}
    \ket{0}_j= \ket{\phi_3}_j\,.
\end{equation}
On such a state, the action of the Lax operator results in a triangular matrix in the auxiliary space,
\begin{equation}
\label{eq:Ltriangular}
\mathcal L_{j,A}(\lambda,p_j) \ket{0}_j = \begin{pmatrix}
\omega_0(\lambda,p_j) \ket{0}_j & * & * & * \\
0 & \omega_1(\lambda,p_j) \ket{0}_j & 0 & * \\
0 & 0 & \omega_2(\lambda,p_j) \ket{0}_j  & * \\
0 & 0 & 0 &\omega_3(\lambda,p_j) \ket{0}_j \\
\end{pmatrix}\,,
\end{equation}
where the stars $*$ denote coefficient which we leave unspecified and the coefficients $\omega_i$ can be expressed in terms of the S-matrix elements:
\begin{equation}
\begin{aligned}
\omega_0(\lambda,p_j) &= A(\lambda,p_j)\,,\qquad &&\omega_1(\lambda,p_j)=L(\lambda,p_j)\,,\\
\omega_2(\lambda,p_j)&=L(\lambda,p_j)\,, \qquad&& \omega_3(\lambda,p_j)=\frac{ A(\lambda,p_j)-B(\lambda,p_j)}{q+q^{-1}}\,.
\end{aligned}
\end{equation}
For a chain with $K^\mathrm{I}$ sites the complete vacuum state is the $K^\mathrm{I}$-fold tensor product
\begin{equation}
\label{eq:vacuum}
\ket{0} = \bigotimes_{j=1}^{K^\mathrm{I}} \ket{0}_j\,.
\end{equation}
We then construct the monodromy matrix
\begin{equation}
\mathcal T_A(\lambda,\vec{p}) = \mathcal L_{K^\mathrm{I},A}(\lambda,p_{K^\mathrm{I}}) \cdots \mathcal L_{1,A}(\lambda,p_1)\,.
\end{equation}
In the auxiliary space, the monodromy matrix is a $4\times4$ matrix whose entries are operators in the physical space, so that we may write, omitting the dependence on $\vec{p}=(p_1,\dots p_{K^\mathrm{I}})$,
\begin{equation}
\mathcal T_A(\lambda) = \begin{pmatrix}
T_0{}^0(\lambda) & R_{0}{}^{\beta}(\lambda) & R(\lambda) \\
S_{\alpha}{}^{0}(\lambda) &
T_{\alpha}{}^{\beta}(\lambda) & R_{\alpha}{}^{3}(\lambda) \\
S(\lambda) & S_{3}{}^{\beta}(\lambda) & T_3{}^3(\lambda)
\end{pmatrix}\,,
\end{equation}
where ${R}_{0}{}^{\beta}(\lambda)$ and $S_{3}{}^{\beta}(\lambda)$ are two-component row vectors, $R_{\alpha}{}^{3}(\lambda)$ and $S_{\alpha}{}^{0}(\lambda)$ are two-component line vectors and $T_{\alpha}{}^{\beta}(\lambda)$ is a $2\times2$ matrix, with $\alpha,\beta \in \{1,2\}$.
The triangular structure~\eqref{eq:Ltriangular} of the Lax operator when acting on the vacuum state allows us to easily determine the action of some of the entries on the vacuum~\eqref{eq:vacuum}. Namely, for the diagonal elements we get
\begin{equation}
\begin{aligned}
&T_0{}^0(\lambda) \ket{0} = \prod_{j=1}^{K^\mathrm{I}} \omega_0(\lambda,\vec{p})\,, \qquad
&&T_1{}^1(\lambda) \ket{0} =  \prod_{j=1}^{K^\mathrm{I}} \omega_1(\lambda,\vec{p})\,,\\
&T_2{}^2(\lambda) \ket{0} =  \prod_{j=1}^{K^\mathrm{I}} \omega_1(\lambda,\vec{p})\,,\qquad
 &&T_3{}^3(\lambda) \ket{0} = \prod_{j=1}^{K^\mathrm{I}} \omega_3(\lambda,\vec{p})\,.
\end{aligned}
\end{equation}
 By virtue of the quantum Yang-Baxter equation~\eqref{eq:YBE} we have the identity
\begin{equation}
\label{eq:RTT}
\check{\mathcal{R}}_{12}(\lambda,\mu)\; \mathcal T(\lambda) \hat{\otimes} \mathcal T(\mu) = \mathcal T(\mu) \hat{\otimes} \mathcal T(\lambda)  \;\check{\mathcal{R}}_{12}(\lambda,\mu)\,, \qquad \check{\mathcal{R}}_{12}= \Pi^g_{12} \mathcal{R}_{12}\,.
\end{equation}
We introduced the graded tensor product~$\hat{\otimes}$ which between two matrices $A_{a}{}^{c}$ and $B_{b}{}^{d}$ is~\cite{Kulish:1980ii}
\begin{equation}
(A \hat{\otimes} B)_{ab}^{cd} = (-1)^{\epsilon_b(\epsilon_a+\epsilon_c)} A_{a}{}^{c} B_{b}{}^{d}\,,
\end{equation}
with $\epsilon_j=0$ for Bosonic excitations and  $\epsilon_j=1$ for Fermionic ones.
From this we can read out the commutation relations for the various operators. 

\subsubsection{Commutation relations for the monodromy matrix}
\label{sec:ABA:BK:commutation}
By spelling out the RTT relation~\eqref{eq:RTT} in components, we get a set of commutation relations among the various matrix elements of the monodromy. Here we write down those which we will need later. For simplicity, we omit the arguments of the functions $A,B,\dots K,L$, which all depend on the auxiliary momenta $\lambda$ and $\mu$ (\textit{i.e.}, $A=A(\lambda,\mu)$, and so on). Furthermore, we have used the following identity to somewhat simplify the expressions\,.
\begin{equation}
\label{eq:relation}
AD = BE-CF = HK-GL\,.
\end{equation}
We find
\begin{equation} \begin{aligned}
{T}_{\alpha}{}^{\beta}(\lambda) {R}_{0}{}^{\gamma}(\mu) &= -\frac{D}{L} \gen{r}_{\rho\sigma}^{\beta\gamma} {R}_{0}{}^{\sigma}(\mu) {T}_{\alpha}{}^{\rho}(\lambda) +\frac{H}{L}  {R}_{0}{}^{\beta}(\lambda)  \hat{T}_{\alpha}{}^{\gamma}(\mu) \\
&\quad - v^{\beta\gamma}\frac{F}{A-B} \left({R}_{\alpha}{}^{3}(\lambda) T_{0}{}^{0}(\mu) + \frac{H}{L} R(\lambda) {S}_{\alpha}{}^{0}(\mu) - \frac{A}{L} R(\mu) {S}_{\alpha}{}^{0}(\lambda) \right)\,,
\end{aligned} 
\end{equation}
as well as
\begin{equation}
T_{0}{}^{0}(\lambda) {R}_{0}{}^{\alpha}(\mu) = - \frac{D}{L} {R}_{0}{}^{\alpha}(\mu) T_{0}{}^{0}(\lambda)+\frac{H}{L} {R}_{0}{}^{\alpha}(\lambda)T_{0}{}^{0}(\mu)\,,
\end{equation}
and
\begin{equation} \begin{aligned}
T_{3}{}^{3}(\lambda) {R}_{0}{}^{\alpha}(\mu) &=   \frac{(q+q^{-1})G}{A-B} {R}_{0}{}^{\alpha}(\mu) T_{3}{}^{3}(\lambda) - \frac{F}{A-B} v^{\beta\gamma}{R}_{\beta}{}^{3}(\lambda) {T}_{\gamma}{}^{\alpha}(\mu)\\
&\qquad-\frac{qA+q^{-1} B}{A-B} R(\lambda) S_{3}{}^\alpha(\mu) + \frac{(q+q^{-1})H}{A-B}  R(\mu) S_{3}{}^\alpha(\lambda)~.
\end{aligned}
\end{equation}
For the abelian raising operator $R(\mu)$ we have
\begin{equation} \begin{aligned}
T_{\alpha}{}^{\beta}(\lambda) R(\mu) &= \left(1+\frac{HK}{GL}\right) R(\mu) T_{\alpha}{}^{\beta}(\lambda) - \frac{H K}{GL}  R(\lambda) T_{\alpha}{}^{\beta}(\mu) \\
&\qquad \qquad - \frac{K}{L}{R}_0{}^\beta(\mu) {R}_\alpha{}^0(\lambda) -\frac{H}{G} {R}_\alpha{}^0(\mu) {R}_0{}^\beta(\lambda)~,
\end{aligned} 
\end{equation}
together with
\begin{equation} \begin{aligned}
 T_0{}^0(\lambda)R(\mu) &= \frac{D(q+q^{-1})}{D-E} R(\mu)  T_0{}^0(\lambda) - \frac{q^{-1} D+ q E}{D-E} R(\lambda)  T_0{}^0(\mu) \\
 &\qquad + \frac{q C}{D-E} v_{\alpha\beta}{R}_0{}^\alpha(\lambda) {R}_0{}^\beta(\mu)\,,
 \end{aligned}
\end{equation}
and
\begin{equation} \begin{aligned}
 T_{3}{}^{3}(\lambda)R(\mu) &= \frac{A(q+q^{-1})}{A-B}R(\mu) T_{3}{}^{3}(\lambda) -  \frac{q A + q^{-1} B}{A-B} R(\lambda) T_{3}{}^{3}(\mu) \\
 &\qquad - \frac{F}{A-B}  v^{\alpha\beta}R_{\alpha}{}^3(\lambda) R_{\beta}{}^3(\mu)\,.
 \end{aligned}
\end{equation}
In the above commutation relations we have introduced an auxiliary S~matrix $\gen{r}$, which is given by
\begin{equation}
\begin{aligned}
&\gen{r}_{11}^{11}= \gen{r}_{22}^{22}=1\,,\qquad
&&\gen{r}_{12}^{12}= \gen{r}_{21}^{21}= b\,, \\
&\gen{r}_{21}^{12}=a\,,\qquad&&\gen{r}_{12}^{21}=c\,,
 \end{aligned}
\end{equation}
where the coefficients are
\begin{equation}
\begin{aligned}
&a=\frac{q D + q^{-1} E}{D(q+q^{-1})} + \frac{q^{-1} C F}{D(A-B)(q+q^{-1})}\,, \\
&b =\frac{D - E}{D(q+q^{-1})} -\frac{C F}{D(A-B)(q+q^{-1})}\,,\\
&c =\frac{q^{-1} D + q E}{D(q+q^{-1})} + \frac{q C F}{D(A-B)(q+q^{-1})}\,.
\end{aligned}
\end{equation}
Furthermore we introduced the vector $v^{\alpha\beta}$ given by
\begin{equation}
    (v^{11},v^{12},v^{21},v^{22})=(v_{11},v_{12},v_{21},v_{22}) = (0,-q^{-1},1,0)\,.
\end{equation}

Let us comment a little more on the auxiliary S~matrix $\gen{r}$. We have the identities
\begin{equation}
a + q^{-1} \,b=1\,, \qquad c + q \,b=1\,,
\end{equation}
so that we can write
\begin{equation} \begin{aligned}
\gen{r}_{11}^{11} &= \gen{r}_{22}^{22}= M~, &\qquad \gen{r}_{12}^{12} &= \gen{r}_{21}^{21}= \frac{M-N}{q+q^{-1}}~, \\
 \gen{r}^{12}_{21} &= \frac{q M + q^{-1} N}{q+q^{-1}}~, &\qquad \gen{r}^{21}_{12} &=  \frac{q^{-1} M + q N}{q+q^{-1}}\,,
\end{aligned}
\end{equation}
with 
\begin{equation}
 M=1~, \qquad N= ac - b^2=- \frac{A}{D}\frac{D-E}{A-B}\,.
\end{equation}
From this expression it is easy to check that the auxiliary S matrix satisfies the quantum Yang-Baxter equation and has $\su_q(2)$ symmetry. Let us go one step further and define the variable
\begin{equation}
\label{eq:defuy}
y_j = x_j^-~, \qquad u_j= y_j + \frac{1}{y_j} = q^{-2} \left( x^+_j + \frac{1}{x^+_j} \right)+ (q^{-2}-1)(\xi+\xi^{-1})\,.
\end{equation}
Then we can write
\begin{equation}
N(u_j,u_k) = \frac{q^{-1} u_j- q u_k -i/\hat{g}}{q^{-1} u_k - q u_j -i/\hat{g}}~, \qquad \hat{g} = \frac{i}{(q-q^{-1})(\xi+\xi^{-1})}\,,
\end{equation}
which is the customary form for an $\su(2)_q$-invariant integrable S~matrix, see \textit{e.g.}~\cite{Beisert:2008tw}.

\subsubsection{Eigenvalue problem}
\label{sec:ABA:BK:eigenvalue}
The eigenvalue problem which we need to solve to find the Bethe equations is given by
\begin{equation}
\text{STr}_A\left[\mathcal T_A(\lambda)\right] \ket{\Phi} = \left(T_{0}{}^{0}(\lambda) - \sum_{\alpha=1}^2 {T}_{\alpha}{}^{\alpha}(\lambda) + T_{3}{}^{3}(\lambda) \right) \ket{\Phi} = \Lambda(\lambda) \ket{\Phi}\,.
\end{equation}
We have already seen that the vacuum~$\ket{0}$ is one of the possible eigenvectors.
As usual~\cite{Faddeev:1996iy}, we can construct more general eigenvectors by acting one or more times on the vacuum by suitable creation operators from the upper-triangular part of $\mathcal{T}_A(\lambda)$. The simplest states are those where we act only once on the vacuum; we call them one-particle states.

\paragraph{One-particle states.}
Introducing a coefficient vector~$X_\alpha$ we make the ansatz
\begin{equation}
\ket{\Phi(u_1)} = R_{0}{}^{\alpha}(u_1)\,X_\alpha \ket{0}\,,
\end{equation}
\textit{i.e.}\ we take the one-particle state to be a yet-to-be-determined combination of ${R}_{0}{}^1(u_1)\ket{0}$ and~${R}_{0}{}^2(u_1)\ket{0}$.
For the operator $T_{0}{}^0(\lambda)$ we have
\begin{align}
T_{0}{}^{0}(\lambda) \ket{\Phi(u_1)} &=X_{\alpha} T_{0}{}^{0}(\lambda) R_{0}{}^{\alpha}(u_1) \ket{0} \\
&= X_{\alpha} \left(- \frac{D}{L} R_{0}{}^{\alpha}(u_1) T_{0}{}^{0}(\lambda) +\frac{H}{L} R_{0}{}^{\alpha}(\lambda)T_{0}{}^{0}(u_1) \right)\ket{0} \\
&= -\frac{D}{L} \Omega_0(\lambda) \ket{\Phi(u_1)} + \frac{H}{L}  \Omega_0(u_1)  R_{0}{}^{\alpha}(\lambda) X_\alpha \ket{0} \,,
\end{align}
where we introduced the short-hand notation
\begin{equation}
\Omega_n(\lambda) = \prod_{j=1}^{K^\mathrm{I}} \omega_n(\lambda,p_j)\,,\qquad n=0,\dots 3\,.
\end{equation}
For the operator $T_{3}{}^3(\lambda)$ we have
\begin{equation}
\begin{aligned}
T_{3}{}^{3}(\lambda) \ket{\Phi(u_1)} &=X_\alpha T_{3}{}^{3}(\lambda) R_{0}{}^{\alpha}(u_1) \ket{0} \\
&= X_\alpha \frac{(q+q^{-1})G}{A-B} R_{0}{}^\alpha(u_1) T_{3}{}^3(\lambda) \ket{0} - X_\alpha\frac{F}{A-B} v^{\beta\gamma} R_{\beta}{}^3(\lambda) T_{\gamma}{}^\alpha(u_1) \\
&= \frac{(q+q^{-1})G}{A-B} \Omega_3(\lambda) \ket{\Phi(u_1)}-\frac{F}{A-B} \Omega_1(u_1)  v^{\beta\gamma} R_\beta{}^3(\lambda) X_\gamma\ket{0}
\end{aligned}
\end{equation}
For the matrix elements ${T}_{\alpha}{}^{\alpha}(\lambda)$ (summing over the indices) we have
\begin{equation} \begin{aligned}
{T}_{\alpha}{}^{\alpha}(\lambda) \ket{\Phi(u_1)} &= {T}_{\alpha}{}^{\alpha}(\lambda) R_{0}{}^\beta(u_1) X_\beta \ket{0} \\
&=  -\frac{F}{A-B} \Omega_0(u_1) v^{\alpha\beta} R_{\alpha}{}^3(\lambda) X_\beta \ket{0} + \frac{H}{L} \Omega_1(u_1) R_{0}{}^\alpha(\lambda) X_{\alpha} \ket{0} \\
&\qquad -\frac{D}{L} \Omega_1(\lambda) \gen{r}_{\alpha\beta}^{\alpha\gamma}  R_0{}^{\beta}(u_1) X_{\gamma} \ket{0}\,.
\end{aligned}
\end{equation}
The terms that are not eigenvectors must cancel. This gives rise to the Bethe equation
\begin{equation}
1=\frac{\Omega_0(u_1)}{\Omega_1(u_1)} = \prod_{j=1}^{K^\mathrm{I}} \frac{\omega_0(u_1,p_j)}{\omega_1(u_1,p_j)}\,.
\end{equation}
The eigenvalue is given by
\begin{equation}
\Lambda(\lambda,\vec{u})=-\frac{D}{L} \Omega_0(\lambda) + \frac{(q+q^{-1}) G}{A-B}\Omega_3(\lambda) - \frac{D}{L} \Omega_1(\lambda) \Lambda^{(1)}(\lambda,u_1)
\end{equation}
where we need to solve the auxiliary problem
\begin{equation}
\gen{r}_{\alpha \gamma}^{\alpha\beta}  X_\beta = \Lambda^{(1)}(\lambda,u_1) X_\gamma\,.
\end{equation}
It turns out that $\gen{r}_{\alpha \gamma}^{\alpha\beta}$ is already diagonal and $\Lambda^{(1)}(\lambda,u_1) = 1+ b(\lambda,u_1)$.

\paragraph{Two-particle states.}
For the two-particle states we make the ansatz
\begin{equation}
\ket{\Phi(u_1,u_2)} = {\Phi}^{\alpha\beta}(u_1,u_2) X_{\alpha\beta} \ket{0}~,
\end{equation}
where
\begin{equation}
{\Phi}^{\alpha\beta}(u_1,u_2) = R_{0}{}^\alpha(u_1) R_{0}{}^\beta(u_2) +  z(u_1,u_2) \, v^{\alpha\beta}  R(u_1) T_{0}{}^{0}(u_2)\,.
\end{equation}
Acting with the diagonal operators will generate two types of terms: the eigenvalue contribution and the unwanted terms. Requiring that the latter vanish will allow us to obtain the function $z(u_1,u_2)$ and write down the Bethe equations. Let us first consider unwanted terms that only arise from acting with one diagonal operator. This is in particular the case for the unwanted term of the form
\begin{equation}
v^{\alpha\beta} R_{\alpha}{}^3(\lambda)  R_{\beta}{}^3(u_1)\,,
\end{equation}
which only appears when applying $T_3{}^3(\lambda)$. Requiring that it vanishes imposes
\begin{equation}
\label{eq:ztwopart}
z(u_1,u_2) = \frac{ F(u_1,u_2)}{A(u_1,u_2)-B(u_1,u_2)}\,.
\end{equation}
With this choice of function one can show that 
\begin{equation}
\label{eq:state_sym}
\Phi^{\alpha\beta} (u_1,u_2) = \frac{D(u_1,u_2)}{A(u_1,u_2)} \Phi^{\delta \gamma}(u_2,u_1) \gen{r}_{\gamma \delta }^{\alpha\beta}(u_1,u_2)~.
\end{equation}
This identity will help us in writing down the unwanted terms, since it is sometimes easier to use the right-hand side rather than the left-hand side of the above equation. In doing so, the following ordering factor will appear:
\begin{equation}
\mathbf{o}(\vec{u})= \frac{D(u_1,u_2)}{A(u_1,u_2)}\,\gen{r}(u_1,u_2)\,.
\end{equation} 
For the diagonal operator $T_0{}^0(\lambda)$ we obtain
\begin{equation} \begin{aligned}
T_0{}^0(\lambda) \ket{\Phi(u_1,u_2)} 
&=  \Omega_0(\lambda) \prod_{j=1}^2 \frac{-D(\lambda,u_j)}{L(\lambda,u_j)}  \ket{\Phi_2(u_1,u_2)}  - \sum_{j=1}^2 \Omega_0(u_j) \ket{\Psi_1(\lambda,u_j)}\\
&\qquad + \Omega_0(u_1) \Omega_0(u_2) H_0(\lambda,u_1,u_2) \ket{\Psi_3(\lambda)}\,,
\end{aligned}
\end{equation}
in terms of some states and functions which we will define below.
For the operator $T_3{}^3(\lambda)$ we have
\begin{equation} \begin{aligned}
T_3{}^3(\lambda) \ket{\Phi(u_1,u_2)} 
&= \Omega_3(\lambda) \prod_{j=1}^2 \frac{(q+q^{-1})G(\lambda,u_j)}{A(\lambda,u_j)-B(\lambda,u_j)}  \ket{\Phi(u_1,u_2)} \\
&\qquad + \sum_{j=1}^2 \Omega_1(u_j) \Lambda^{(1)}(u_j)\ket{\Psi_2(\lambda,u_j)} \\
&\qquad + \Omega_1(u_1) \Omega_1(u_2) H_1(\lambda,u_1,u_2)  \ket{\Psi_3(\lambda)}~.
\end{aligned}
\end{equation}

Finally, for the operators $T_\alpha{}^\alpha(\lambda)$ we get
\begin{equation} \begin{aligned}
T_\alpha{}^\alpha(\lambda) \ket{\Phi(u_1,u_2)}  &= \Omega_1(\lambda) \prod_{j=1}^2 \frac{-D(\lambda,u_j)}{L(\lambda,u_j)}\Lambda^{(1)}(\lambda,u_1,u_2) \ket{\Phi(u_1,u_2)}  \\
&\quad - \sum_{j=1}^2 \Omega_1(u_j) \Lambda^{(1)}(u_j)\ket{\Psi_1(\lambda,u_j)}
+ \sum_{j=1}^2 \Omega_0(u_j) \ket{\Psi_2(\lambda,u_j)} \\
&\quad -  \Omega_0(u_1) \Omega_1(u_2) \Lambda^{(1)}(u_2) H_2(\lambda,u_1,u_2) \ket{\Psi_3(\lambda)} \\
&\quad - \Omega_0(u_2) \Omega_1(u_1) \Lambda^{(1)}(u_1) H_3(\lambda,u_1,u_2) \ket{\Psi_3(\lambda)}~.
\end{aligned}
\end{equation}
In the above we denoted by $\Lambda^{(1)}(\lambda)$ the eigenvalue of the auxiliary problem
\begin{equation}
\gen{r}^{\alpha \gamma}_{\beta \rho}(\lambda,u_1)\,  \gen{r}^{\beta \delta}_{\alpha \sigma}(\lambda,u_2)\, X_{\gamma\delta} = \Lambda^{(1)}(\lambda)\,  X_{\rho\sigma}\,.
\end{equation}
We will come back to this auxiliary problem in the next section. We gathered the three types of unwanted terms into
\begin{equation}
\begin{aligned} 
\ket{\Psi_1(\lambda,u_j)} &= \frac{H(\lambda,u_j)}{L(\lambda,u_j)} \prod_{k \neq j}^2 \frac{D(u_j,u_k)}{L(u_j,u_k)} R_{0}{}^\alpha(\lambda) R_{0}{}^\beta(u_k) (\delta_j^1\delta_\alpha^\gamma \delta_\beta^\delta + \delta_j^2 \gen{o}^{\gamma\delta}_{\beta \alpha}(\vec{u})) X_{\gamma\delta} \ket{0}~, \\
\ket{\Psi_2(\lambda,u_j)} &=\frac{ F(\lambda,u_j)}{A(\lambda,u_j)-B(\lambda,u_j)} \prod_{k \neq j}^2\frac{D(u_j,u_k)}{L(u_j,u_k)}
v^{\rho\alpha}R_{\rho}{}^3(\lambda) R_0{}^\beta(u_k) (\delta_j^1\delta_\alpha^\gamma \delta_\beta^\delta + \delta_j^2 \gen{o}^{\gamma\delta}_{\beta \alpha}(\vec{u})) X_{\gamma\delta} \ket{0}\,, \\
\ket{\Psi_3(\lambda)} &= R(\lambda) v^{\alpha\beta} X_{\alpha\beta} \ket{0}~,
\end{aligned}
\end{equation}
and introduced the functions
\begingroup
\allowdisplaybreaks
\begin{align} 
\nonumber
H_1(\lambda,u_1,u_2) &= -\frac{q^{-1} D(\lambda,u_1)+q E(\lambda,u_1)}{D(\lambda,u_1)-E(\lambda,u_1)} \frac{ F(u_1,u_2)}{A(u_1,u_2)-B(u_1,u_2)}\\
&\qquad\qquad\qquad\qquad
+  \frac{D(\lambda,u_1)}{L(\lambda,u_1)} \frac{H(\lambda,u_2)}{L(\lambda,u_2)}\frac{ F(\lambda,u_1)}{D(\lambda,u_1)-E(\lambda,u_1)}\,, \\
\nonumber
H_2(\lambda,u_1,u_2) &= -\frac{q A(\lambda,u_1)+q^{-1} B(\lambda,u_1)}{A(\lambda,u_1)-B(\lambda,u_1)} \frac{ F(u_1,u_2)}{A(u_1,u_2)-B(u_1,u_2)}\\
&\qquad\qquad\qquad\qquad +  \frac{(q+q^{-1})H(\lambda,u_1)}{A(\lambda,u_2)-B(\lambda,u_2)}\frac{ F(\lambda,u_2)}{A(\lambda,u_2)-B(\lambda,u_2)}\,, \\
\nonumber
H_3(\lambda,u_1,u_2) &= \left(\frac{A(\lambda,u_1) H(\lambda,u_2)}{L(\lambda,u_1) L(\lambda,u_2)} - \frac{H(u_1,u_2) H(\lambda,u_1)}{L(u_1,u_2) L(\lambda,u_1)}\right)\\
&\qquad\qquad\qquad\qquad\times \frac{ F(\lambda,u_1)}{A(\lambda,u_1)-B(\lambda,u_1)} N(u_1,u_2)\,, \\
\nonumber
H_4(\lambda,u_1,u_2) &= H_3(\lambda,u_2,u_1) \frac{1}{N(u_2,u_1)} \frac{D(u_1,u_2)}{A(u_1,u_2)} \\
\nonumber
&= \left(\frac{A(\lambda,u_2) H(\lambda,u_1)}{L(\lambda,u_2) L(\lambda,u_1)} + \frac{K(u_1,u_2) H(\lambda,u_2)}{L(u_1,u_2) L(\lambda,u_2)}\right)\\
&\qquad\qquad\qquad\qquad\times \frac{ F(\lambda,u_2)}{A(\lambda,u_2)-B(\lambda,u_2)}\frac{D(u_1,u_2)}{A(u_1,u_2)}~.
\end{align}
\endgroup

We observe that in order for the unwanted terms proportional to $\ket{\Psi_1(\lambda,u_j)}$ and $\ket{\Psi_2(\lambda,u_j)}$ to cancel, we need to impose the Bethe equations
\begin{equation}
\frac{\Omega_0(u_j)}{\Omega_1(u_j)} =  \Lambda^{(1)}(u_j)\,, \qquad j=1,2\,.
\end{equation}
We are then left with showing that the unwanted terms proportional to $\ket{\Psi_3(\lambda)}$ cancel. To achieve this we use the Bethe equation to factor out a common $\Omega_2(u_1)\Omega_2(u_2)$, and use the identity
\begin{equation}
\Lambda^{(1)}(u_1) \Lambda^{(1)}(u_2)=1
\,.
\end{equation}
One then arrives at the condition
\begin{equation}
H_1 + H_2 + H_3 + H_4 =0~,
\end{equation}
which we verified numerically.

Obtaining the Bethe equations by examining the unwanted terms is still possible in the case of two excitations, but the complexity of the computations increases significantly as we include more and more excitations.
There is however another way to obtain the Bethe equations, by requiring that the eigenvalue of the transfer matrix is regular. This is the approach we will take to obtain the Bethe equations for the multi-particle state.

\paragraph{The $K^\mathrm{II}$-particle states.}
For a $K^\mathrm{II}$-particle state we assume that the eigenvector takes the form
\begin{equation}
\ket{\Phi(u_1,\dots u_{\ind{K^\mathrm{II}}})} = {\Phi}^{\alpha_1\dots \alpha_{\ind{K^\mathrm{II}}}}(u_1,\dots, u_{\ind{K^\mathrm{II}}}) X_{\alpha_1\dots \alpha_{K^\mathrm{II}}} \ket{0}~,
\end{equation}
where the wavefunction is defined by the recursive relation
\begin{equation}
\label{eq:nparticleansatz}
\begin{aligned}
{\Phi}^{\alpha_1\dots \alpha_{K^\mathrm{II}}}(u_1,\dots, u_{K^\mathrm{II}}) &= {R}_0{}^{\alpha_1}(u_1) {\Phi}^{\alpha_2\dots\alpha_{K^\mathrm{II}}}(u_2,\dots, u_{K^\mathrm{II}})\\
&\quad +\sum_{j=2}^{K^\mathrm{II}} v^{\alpha_1\alpha_{j}}R(u_1) {\Phi}_{(j)}^{\alpha_2\dots\alpha_{K^\mathrm{II}}}(u_2,\dots,u_{K^\mathrm{II}}) T_0{}^0(u_j) z_{(j)}(u_1,\dots u_{K^\mathrm{II}})\,,
\end{aligned}
\end{equation}
where ${\Phi}_{(j)}^{\alpha_2\dots\alpha_{K^\mathrm{II}}}(u_2,\dots,u_{K^\mathrm{II}})$ depends on $(K^\mathrm{II}-2)$ particles only, as it does not depend on $u_j$ and $\alpha_j$. By convention we take ${\Phi} = 1$ for no excitation. Here $z_{(j)}(u_1,\dots u_{K^\mathrm{II}})$ is a generalisation of~\eqref{eq:ztwopart} but it will not be necessary to work out its form for our purposes. As we mentioned, we are only interested in computing the putative eigenvalue (assuming that an eigenvector exists) and find the Bethe equations by requiring that it is regular. It turns out that it is sufficient to work with the first line of eq.~\eqref{eq:nparticleansatz} to determine the eigenvalue. In fact, looking at the commutation relations we seen that the action of any $T_{A}{}^{A}(\lambda)$ on the second line cannot generate a term involving $\prod_j R_{0}{}^{\alpha_j}(u_j)$, because, loosely speaking, one of the $R_{0}{}^{\alpha_j}(u_j)$ is missing---namely, the one appearing in $T_{0}{}^{0}(u_j)$ in the second line of eq.~\eqref{eq:nparticleansatz}.
Therefore, let us now compute the eigenvalue of the transfer matrix $T_A{}^A(\lambda)$ by looking at the first line of eq.~\eqref{eq:nparticleansatz}. We need only one term in the commutation relations, which greatly simplifies the computation, yielding the eigenvalue $\Lambda(\lambda,\vec{\mu}_j)$
\begin{equation}
\begin{aligned}
\Lambda(\lambda,\vec{\mu}_j)&=
 \Omega_0(\lambda) \prod_{j=1}^{K^\mathrm{II}}  \frac{-D(\lambda,u_j)}{L(\lambda,u_j)} + \Omega_3(\lambda) \prod_{j=1}^{K^\mathrm{II}} \frac{(q+q^{-1})G(\lambda,u_j)}{A(\lambda,u_j)-B(\lambda,u_j)}\\
 &\quad- \Omega_1(\lambda) \prod_{j=1}^{K^\mathrm{II}} \frac{-D(\lambda,u_j)}{L(\lambda,u_j)}\Lambda^{(1)}(\lambda,\vec{u})\,,
\end{aligned}
\end{equation}
where $\Lambda^{(1)}(\lambda,\vec{\mu})$ is the solution of the auxiliary eigenvalue problem
\begin{equation}
\label{eq:auxiliaryproblem}
\gen{r}^{\alpha\beta_1}_{\gamma_1 \delta_1}(\lambda,u_1) \gen{r}^{\gamma_1  \beta_2}_{\gamma_2 \delta_2}(\lambda,u_2) \cdots \gen{r}^{\gamma_{K^\mathrm{II}-1}  \beta_{K^\mathrm{II}}}_{\alpha \delta_{K^\mathrm{II}}}(\lambda,u_{K^\mathrm{II}})\, X_{\beta_1 \dots \beta_{K^\mathrm{II}}} = \Lambda^{(1)}(\lambda,\vec{u})\, X_{\delta_1 \dots \delta_{K^\mathrm{II}}}\,,
\end{equation}
which we shall solve in the next section.
Let us now consider the regularity of the eigenvalues. The function $L(\lambda,u_j)$ has a zero when $\lambda=u_j$; $D(\lambda,u_j)=-1$ is constant hence regular. Both $G(\lambda,u_j)$ and $A(\lambda,u_j)-B(\lambda,u_j)$ have zeros when $\lambda=u_j$, but they cancel each other out and the ratio is regular.  Requiring that the residue of $\Lambda(\lambda,\vec{u})$ in $\lambda=u_j$ vanishes imposes the constraint
\begin{equation}
\label{eq:Bethe1}
\frac{\Omega_0(u_j)}{\Omega_1(u_j)} =  \Lambda^{(1)}(u_j,\vec{u})\,,
\end{equation}
which we should impose for $j=1,\dots K^\mathrm{II}$, \textit{i.e.}\ for all rapidities.

\subsubsection{Auxiliary eigenvalue problem}
\label{sec:ABA:BK:auxiliary}
For the $K^\mathrm{II}$-particle state the auxiliary problem reads as in eq.~\eqref{eq:auxiliaryproblem}.
This involves product of $\su(2)_q$-invariant S~matrices with inhomogeneities $\{u_1,\dots u_{K^\mathrm{II}}\}$, acting on $X_{\beta_1\dots\beta_{K^\mathrm{II}}}$. The auxiliary S~matrix $\gen{r}$ satisfies the quantum Yang-Baxter equation and can itself be diagonalised by means of the algebraic Bethe ansatz. The auxiliary space is two-dimensional $\mathcal H_A^{(1)} = \mathbb{C}^2$ and the Lax operators read
\begin{equation}
\mathcal L^{(1)}_{j,A} (\lambda,u_j)=  \gen{r}_{A,j}(\lambda,u_j)\,.
\end{equation}
From these we construct the monodromy matrix
\begin{equation}
\mathcal T^{(1)}(\lambda,\{u_j\}) = \mathcal L_{K^\mathrm{II},A}^{(1)}(\lambda,u_{K^\mathrm{II}}) \cdots \mathcal L^{(1)}_{1,A}(\lambda,u_1)\,.
\end{equation}
We choose as reference state 
\begin{equation}
\label{eq:basisauxiliary}
\ket{0}^{(1)} = \bigotimes_{j=1}^{K^\mathrm{II}}\ket{0}_j\,,\qquad
\ket{0}\cong\begin{pmatrix} 1 \\ 0 \end{pmatrix}\,,
\end{equation} 
and make the ansatz
\begin{equation}
\mathcal T^{(1)}(\lambda,\vec{u}) = \begin{pmatrix}
A^{(1)}(\lambda,\vec{u}) & B^{(1)}(\lambda,\vec{u}) \\
C^{(1)}(\lambda,\vec{u}) & D^{(1)}(\lambda,\vec{u})
\end{pmatrix}\,,
\end{equation}
so that
\begin{align}
A^{(1)}(\lambda,\vec{u}) \ket{0}^{(1)} &= \ket{0}^{(1)}\,, \\
D^{(1)}(\lambda,\vec{u}) \ket{0}^{(1)} &= \prod_{j=1}^{K^\mathrm{II}} b(\lambda,u_j) \ket{0}^{(1)}\,, \\
C^{(1)}(\lambda,\vec{u}) \ket{0}^{(1)} &= 0\,.
\end{align}

\paragraph{Commutation relations.}
By virtue of the quantum Yang-Baxter equation, the monodromy matrix solves the RTT relations,
from which we can read off the commutation relations
\begin{equation}
\begin{aligned}
A^{(1)}(\lambda) B^{(1)}(\mu) &= \frac{b^2-ac}{b} B^{(1)}(\mu) A^{(1)}(\lambda)+\frac{a}{b} B^{(1)}(\lambda) A^{(1)}(\mu)\,,\\
D^{(1)}(\lambda) B^{(1)}(\mu) &= \frac{1}{b} B^{(1)}(\mu) D^{(1)}(\lambda)-\frac{a}{b} B^{(1)}(\lambda) D^{(1)}(\mu) \,.
\end{aligned}
\end{equation}
For a $K^\mathrm{III}$-particle state with auxiliary rapidities $v_1,\dots v_{K^\mathrm{III}}$ we make the ansatz
\begin{equation}
\ket{\Phi(v_1,\dots v_{K^\mathrm{III}})}^{(1)} = B^{(1)}(v_1) \cdots B^{(1)}(v_{K^\mathrm{III}}) \ket{0}^{(1)}~.
\end{equation}
The auxiliary eigenvalue is then
\begin{equation}
\label{eq:auxiliaryeigenvalueabc}
\Lambda^{(1)}(\lambda,\vec{u})= \prod_{k=1}^{K^\mathrm{III}} \frac{b(\lambda,v_k)^2-a(\lambda,v_k) c(\lambda,v_k)}{b(\lambda,v_k)} + \prod_{k=1}^{K^\mathrm{III}} \frac{1}{b(\lambda,v_k)} \prod_{j=1}^{K^\mathrm{II}} b(\lambda,u_j)\,,
\end{equation}
which can also be written as
\begin{equation}
\label{eq:auxiliaryeigenvalue}
\begin{aligned}
\Lambda^{(1)}(\lambda,\vec{u}) &= \prod_{k=1}^{K^\mathrm{III}} \frac{-(q+q^{-1})N(\lambda,v_k)}{1-N(\lambda,v_k)} + \prod_{k=1}^{K^\mathrm{III}} \frac{q+q^{-1}}{1-N(\lambda,v_k)} \prod_{j=1}^{K^\mathrm{II}} \frac{1-N(\lambda,u_j)}{q+q^{-1}}\\
&= \prod_{k=1}^{K^\mathrm{III}} \frac{q^{-1} \lambda-q v_k -i/\hat{g}}{\lambda-v_k} + \prod_{k=1}^{K^\mathrm{III}} \frac{q \lambda - q^{-1} v_k +i/\hat{g}}{\lambda-v_k} \prod_{j=1}^{K^\mathrm{II}} \frac{\lambda-u_j}{q\lambda-q^{-1}u_j+i/\hat{g}}~.
\end{aligned}
\end{equation}

\paragraph{Regularity condition and Bethe equations.}
Let us now require that the eigenvalue $\Lambda^{(1)}(\lambda,\vec{u})$ is regular. There is an apparent pole in~\eqref{eq:auxiliaryeigenvalue} at $\lambda=v_k$, whose residue is proportional to
\begin{equation}
\prod_{l \neq k}^{K^\mathrm{III}} \frac{q^{-1} v_k - q v_l - i/\hat{g}}{v_k-v_l} - \prod_{l \neq k}^{K^\mathrm{III}} \frac{ q v_k - q^{-1} v_l +i/\hat{g}}{v_k - v_l}  \prod_{j=1}^{K^\mathrm{II}} \frac{v_k-u_j}{qv_k-q^{-1}u_j+i/\hat{g}}\,.
\end{equation}
Requiring that it vanishes gives the Bethe equations
\begin{equation}
\prod_{j=1}^{K^\mathrm{II}} \frac{v_k-u_j}{qv_k-q^{-1}u_j+i/\hat{g}}  = \prod_{l \neq k}^{K^\mathrm{III}} \frac{q^{-1} v_k-q v_l-i/\hat{g}}{q v_k -q^{-1} v_l +i/\hat{g}}\,, \qquad k=1,\dots K^\mathrm{III}\,,
\end{equation}
or, in terms of the coefficients $a,b,c$,
\begin{equation}
 \prod_{j=1}^{K^\mathrm{II}} b(v_k,u_j) = \prod_{l \neq k}^{K^\mathrm{III}}  \left[b(v_k,v_l)^2 -a(v_k,v_l) c(v_k,v_l) \right]\,.
\end{equation}

\paragraph{Summary of the Bethe equations.}
It is convenient to introduce
\begin{equation}
R^\mathrm{III,III}(v_j,v_k) =  \frac{q^{-1} v_k-q v_j-i/\hat{g}}{q v_k -q^{-1} v_j+i/\hat{g}}~, \qquad R^\mathrm{II,III}(u_j,v_k) = \frac{q^{-1}u_j-q v_k - i/\hat{g}}{u_j-v_k}\,, 
\end{equation}
so that the auxiliary Bethe equation reads simply
\begin{equation}
1=\prod_{j=1}^{K^\mathrm{II}} R^\mathrm{II,III}(u_j, v_k) \prod_{j \neq k}^{K^\mathrm{III}} R^\mathrm{III,III}(v_j,v_k)\,, \qquad k=1,\dots K^\mathrm{III}\,,
\end{equation}
Furthermore, the main Bethe equation \eqref{eq:Bethe1} can be written as
\begin{equation}
1 = \prod_{j=1}^{K^\mathrm{I}} R^\mathrm{I,II}(x_j,u_k) \prod_{j=1}^{K^\mathrm{III}} R^\mathrm{III,II}(v_j,u_k)\,,  \qquad k=1,\dots K^\mathrm{II}\,,
\end{equation}
with
\begin{equation}
 \qquad R^\mathrm{II,I}(u_k,x_j)= q^{1/2} U_j V_j \frac{y_k - x_j^-}{y_k-x_j^+}\,.
\end{equation}
We remind the reader that we make the identification \eqref{eq:defuy}.
These are precisely the Bethe equations found by Beisert and Koroteev (BK), in the new variables
\begin{equation}
v_k = q^{-1}\left(w_k^{\text{BK}} - \frac{i}{2 g^{\text{BK}}}\right)\,, \qquad \hat{g}=g^{\text{BK}}\,.
\end{equation}

\subsection{Algebraic Bethe ansatz for the Fermionic S~matrix}
\label{sec:ABA:Fermionic}
In this section we analyse how the twist affects the algebraic Bethe ansatz. We will derive the Bethe ansatz for the rescaled fermionic S matrix of section~\ref{sec:Smatrix:Fermionicrescaled}, as this turns out to be simpler. The construction of section~\ref{sec:ABA:BK} can be repeated \textit{verbatim} and the action of the monodromy matrix on the vacuum is the same for the Beisert-Koroteev and for the Fermionic S~matrix. What does change is the form of the commutation relations.

\subsubsection{Commutation relations}
\label{sec:ABA:Fermionic:commutation}

Following the discussion above, we will be interested in obtaining the Bethe equations by demanding that the eigenvalues of the transfer matrix are regular. To this end it is sufficient to consider a few commutation relations; moreover, only one term per equation will actually be relevant for our purpose. We highlight the terms which will play a role by a box. Moreover, we highlight in red the terms which are due to the twist described in section~\ref{sec:Smatrix:twist}. One relevant set of commutation relations is
\begin{equation} \begin{aligned}
&&{T}_{\alpha}{}^{\beta}(\lambda) {R}_{0}{}^{\gamma}(\mu) = \boxed{\frac{-D}{L} {\color{red}\tilde{\gen{r}}_{\rho\sigma}^{\beta\gamma}} {R}_{0}{}^{\sigma}(\mu) {T}_{\alpha}{}^{\rho}(\lambda)} +\frac{H}{L}  {R}_{0}{}^{\beta}(\lambda)  {T}_{\alpha}{}^{\gamma}(\mu)\qquad\qquad \\
&&\qquad - {\color{red}\tilde{w}^{\beta\gamma}}\frac{F}{A-B} \Big({R}_{\alpha}{}^{3}(\lambda) T_{0}{}^{0}(\mu) + \frac{H}{L} R(\lambda) {S}_{\alpha}{}^{0}(\mu)\qquad\\
&&- {\color{red}\tilde{W}_{\alpha}^{\delta}} R(\mu) {S}_{\delta}{}^{0}(\lambda) \Big)\,,
\end{aligned} 
\end{equation}
where the only consequential change concerns the auxiliary S~matrix $\tilde{\gen{r}}$ ($\tilde{w}^{\alpha\beta}$ and $\tilde{W}_\alpha^\beta$ are suitable functions whose form will not be important in what follows). Moreover, we will need two more commutation relations, namely
\begin{equation}
T_{0}{}^{0}(\lambda) {R}_{0}{}^{\alpha}(\mu) =  \boxed{\frac{-D}{L} {R}_{0}{}^{\alpha}(\mu) T_{0}{}^{0}(\lambda)}+\frac{H}{L} {R}_{0}{}^{\alpha}(\lambda)T_{0}{}^{0}(\mu)\,,
\end{equation}
which is completely unchanged, as well as
\begin{equation} \begin{aligned}
T_{3}{}^{3}(\lambda) {R}_{0}{}^{\alpha}(\mu) &=   \boxed{\frac{(q+q^{-1})G}{A-B} {R}_{0}{}^{\alpha}(\mu) T_{3}{}^{3}(\lambda)} - \frac{F}{A-B} {\color{red}\tilde{w}^{\beta\gamma}}{R}_{\beta}{}^{3}(\lambda) {T}_{\gamma}{}^{\alpha}(\mu)\\
&\qquad-{\color{red}\varphi_{12}\frac{g_1}{g_2}}\frac{qA+q^{-1} B}{A-B} R(\lambda) S_{3}{}^\alpha(\mu) + \frac{(q+q^{-1})H}{A-B}  R(\mu) S_{3}{}^\alpha(\lambda)\,.
\end{aligned}
\end{equation}
Even if they will not be needed in what follows, for completeness let us write down the commutation relations involving $R(\mu)$, that are
\begin{equation} \begin{aligned}
T_{\alpha}{}^{\beta}(\lambda) R(\mu) &= \boxed{\left(1+\frac{HK}{GL}\right) R(\mu) T_{\alpha}{}^{\beta}(\lambda)} - {\color{red}\tilde{Z}_{\alpha\gamma}^{\beta\delta}}\frac{H K}{GL}  R(\lambda) T_{\delta}{}^{\gamma}(\mu) \\
&\qquad \qquad  -\frac{H}{G} {R}_\delta{}^0(\mu) {R}_0{}^\beta(\gamma)- {\color{red}\tilde{Z}_{\alpha\gamma}^{\beta\delta}}\frac{K}{L}{R}_0{}^\beta(\mu) {R}_\alpha{}^0(\lambda)\,,
\end{aligned}
\end{equation}
as well as
\begin{equation} \begin{aligned}
 T_0{}^0(\lambda)R(\mu) &= \boxed{\frac{D(q+q^{-1})}{D-E} R(\mu)  T_0{}^0(\lambda)} - {\color{red}\hat{\varphi}_{21}\frac{g_1}{g_2}}\frac{q^{-1} D+ q E}{D-E} R(\lambda)  T_0{}^0(\mu) \\
 &\qquad + \frac{q^2 C}{D-E} {\color{red}\frac{f_2}{g_2}\tilde{v}_{\alpha\beta}(\mu)}{R}_0{}^\alpha(\lambda) {R}_0{}^\beta(\mu)\,,
 \end{aligned}
\end{equation}
and
\begin{equation} \begin{aligned}
 T_{3}{}^{3}(\lambda)R(\mu) &= \boxed{\frac{A(q+q^{-1})}{A-B}R(\mu) T_{3}{}^{3}(\lambda)} -  {\color{red}\varphi_{12} \frac{g_1}{g_2}}\frac{q A + q^{-1} B}{A-B} R(\lambda) T_{3}{}^{3}(\mu) \\
 &\qquad - \frac{q^{-1} F}{A-B}  {\color{red}\tilde{v}^{\alpha\beta}(\lambda)}R_{\alpha}{}^3(\lambda) R_{\beta}{}^3(\mu)\,.
 \end{aligned}
\end{equation}
Without delving to much on the form of the precise form of the various new functions (in red) which we have introduced, we see that the only one which may affect the computation of the eigenvalues of the transfer matrix (and hence the Bethe equations) is the auxiliary S~matrix $\tilde{\gen{r}}$, on which we therefore focus our attention.

\subsubsection{The auxiliary S~matrix and its diagonalisation}
\label{sec:ABA:Fermionic:auxiliary}
The explicit form of the auxiliary S~matrix is now
\begin{equation}
\begin{aligned}
&\tilde{\gen{r}}^{11}_{11}= \tilde{\gen{r}}^{22}_{22}=1\,,\qquad&&
\tilde{\gen{r}}^{12}_{12}= \tilde{\gen{r}}^{21}_{21}= \tilde{b}\,, \\
&\tilde{\gen{r}}^{12}_{21}=\tilde{a}\,,\qquad &&\tilde{\gen{r}}^{21}_{12}=\tilde{c}\,,
 \end{aligned}
\end{equation}
with
\begin{equation}
\begin{aligned}
\tilde{a}&=\varphi_{21} \frac{f_2}{f_1} \frac{q D + q^{-1} E}{D(q+q^{-1})} + \frac{g_1}{f_1}\frac{q^{-1} C F}{D(A-B)(q+q^{-1})}\,, \\
\tilde{b} &=\frac{D - E}{D(q+q^{-1})} -\frac{C F}{D(A-B)(q+q^{-1})}\,,\\
\tilde{c} &= \hat{\varphi}_{21} \frac{f_1}{f_2}\frac{q^{-1} D + q E}{D(q+q^{-1})} + \frac{f_1}{g_1}\frac{q C F}{D(A-B)(q+q^{-1})}\,.
\end{aligned}
\end{equation}
We note the following identities:
\begin{equation}
\label{eq:abcidentities}
\tilde{b}=b\,,\qquad
\tilde{a} \tilde{c} - \tilde{b}^2=ac - {b}^2=N\,.
\end{equation}
Furthermore, the auxiliary S matrix again satisfies the quantum Yang-Baxter equation. In fact, $\tilde{\gen{r}}$ is related to $\gen{r}$ by a change of basis that acts on the basis of $\mathbb{C}^2$ of eq.~\eqref{eq:basisauxiliary} as a rapidity-dependent rescaling. Namely, we set
\begin{equation}
\label{eq:changeofbasis}
    \gen{u}(u)=\begin{pmatrix}
    1 & 0 \\ 0&h(u) \end{pmatrix}\,,\qquad
    \gen{u}(u_1,u_2)=\gen{u}(u_1)\otimes\gen{u}(u_2)\,,
\end{equation}
and observing that
\begin{equation}
\frac{\tilde{a}(u_1,u_2)}{a(u_1,u_2)} = \frac{h(y_1)}{h(y_2)}\,, \qquad \frac{\tilde{c}(u_1,u_2)}{c(u_1,u_2)} = \frac{h(y_2)}{h(y_1)}\,, \qquad h(y) = \frac{y }{y+\xi}\,,
\end{equation}
where we remind that $y$ is related to the rapidity $u$ as
\begin{equation}
    u=y+\frac{1}{y}\,,
\end{equation}
we have that
\begin{equation}
    \gen{u}(u_1,u_2)^{-1}\,\tilde{\gen{r}}(u_1,u_2)\,\gen{u}(u_1,u_2)=\gen{r}(u_1,u_2)\,.
\end{equation}

In conclusion, for the auxiliary S~matrix $\tilde{\gen{r}}$ we can repeat the algebraic Bethe ansatz derivation of section~\ref{sec:ABA:BK:auxiliary}, given that $\tilde{\gen{r}}$ satisfies the Yang-Baxter equation. The eigenvalue will take the same form as eq.~\eqref{eq:auxiliaryeigenvalueabc}, but now expressed in terms of $\tilde{a},\tilde{b},\tilde{c}$:
\begin{equation}
\label{eq:eigtilde}
\tilde{\Lambda}^{(1)}(\lambda,\vec{u})= \prod_{k=1}^{K^\mathrm{III}} \frac{\tilde{b}(\lambda,v_k)^2-\tilde{a}(\lambda,v_k) \tilde{c}(\lambda,v_k)}{\tilde{b}(\lambda,v_k)} + \prod_{k=1}^{K^\mathrm{III}} \frac{1}{\tilde{b}(\lambda,v_k)} \prod_{j=1}^{K^\mathrm{II}} \tilde{b}(\lambda,u_j)\,.
\end{equation}
In view of \eqref{eq:abcidentities} it follows that
\begin{equation}
    \tilde{\Lambda}^{(1)}(\lambda,\vec{u})=\Lambda^{(1)}(\lambda,\vec{u})\,,
\end{equation}
so that \emph{all the eigenvalues of the Fermionic S~matrix coincide with the Beisert-Koroteev ones}. This in particular implies that \emph{the Bethe equations will take the same form too}. As for the eigenvectors, they will be different, and for the auxiliary problem the change of basis from one set of eigenvectors to the other will be given by the matrix $\gen{u}$ of eq.~\eqref{eq:changeofbasis}.
We shall see in appendix~\ref{app:CBA} that similar considerations would apply had we derived the Bethe equations from the coordinate Bethe ansatz.

\section{Bethe-Yang equations for the Fermionic deformation}
\label{sec:result}
The full Fermionic S matrix is given by the tensor product of two $\su(2|2)_q$-invariant Fermionic S matrices, one with deformation parameter $q$ and the other with deformation parameter $q^{-1}$. The ten coefficients $A,B\dots,L$ are invariant under the transformation $q\rightarrow q^{-1}$, while 
\begin{equation}
    f_j \rightarrow \frac{1}{f_j}(1-\xi^2)~, \qquad g_j \rightarrow \frac{1}{g_j}(1-\xi^2)~, \qquad \varphi_{ij} \rightarrow \hat{\varphi}_{ij}~, \qquad \hat{\varphi}_{ij} \rightarrow \varphi_{ij}~.
\end{equation}
It turns out that the spectrum is left invariant under this transformation, as its only effect in the highlighted terms of the commutation relations of section~\ref{sec:ABA:Fermionic:commutation} is to exchange $\tilde{a}$ and $\tilde{c}$ in the auxiliary S matrix. As can be seen from \eqref{eq:eigtilde}, the auxiliary eigenvalue and Bethe equations are not affected by this swapping.

The Bethe-Yang equations for a state of length $J$ then read as follows. There is a main equation that comes from requiring periodicity, 
\begin{equation}
    1 =e^{i J p_k}  \prod_{j\neq k}^{K^\mathrm{I}} S_0(x_j,x_k)   \prod_{\mu=\pm} \prod_{j=1}^{K^\mathrm{II}_{(\mu)}} R_{(\mu)}^\mathrm{II,I}(u_j^{(\mu)},x_k)\,, \qquad k=1,\dots , K^\mathrm{I}\,,
\end{equation}
where $\mu=+,-$ denotes the two copies, and the scattering element $S_0(x_j,x_k)$ is given by
\begin{equation}
    S_0(x_1,x_2) = \Sigma(x_1,x_2)\,A(x_1,x_2)^2\,.
\end{equation}
Here $\Sigma(x_1,x_2)$ is an appropriate dressing factor. Given the the crossing equations are the same~\cite{Seibold:2020ywq} as for the model studied in~\cite{Hoare:2011wr}, the dressing factor proposed there is a natural candidate for~$\Sigma(x_1,x_2)$.
The auxiliary Bethe equations read
\begin{align}
1 &= \prod_{j=1}^{K^\mathrm{I}} R^\mathrm{I,II}_{(\mu)}(x_j,u_k^{(\mu)}) \prod_{j=1}^{K^\mathrm{III}_{(\mu)}} R^\mathrm{III,II}(v_j^{(\mu)},u_k^{(\mu)})\,, \qquad k=1,\dots , K^\mathrm{II}_{(\mu)}\,, \\
1 &=\prod_{j=1}^{K^\mathrm{II}_{(\mu)}} R^\mathrm{II,III}(u_j^{(\mu)}, v_k^{(\mu)}) \prod_{l \neq k}^{K^\mathrm{III}_{(\mu)}} R^\mathrm{III,III}(v_l^{(\mu)},v_k^{(\mu)})\,, \qquad k=1,\dots , K^\mathrm{III}_{(\mu)}\,,
\end{align}
where the auxiliary functions take the same form in both copies,
\begin{align}
R^\mathrm{II,I}_{(\mu)}(u_k^{(\mu)},x_j) &= q^{1/2} U_j V_j \frac{y_k^{(\mu)} - x_j^-}{y_k^{(\mu)}-x_j^+}\,, \\
R^\mathrm{II,III}_{(\mu)}(u_j^{(\mu)},v_k^{(\mu)}) &= \frac{q^{-1}u_j^{(\mu)}-q v_k^{(\mu)} - i/\hat{g}}{u_j^{(\mu)}-v_k^{(\mu)}}\,,  \\
R^\mathrm{III,III}_{(\mu)}(v_j^{(\mu)},v_k^{(\mu)}) &=  \frac{q^{-1} v_k^{(\mu)}-q v_j^{(\mu)}-i/\hat{g}}{q v_k^{(\mu)} -q^{-1} v_j^{(\mu)}+i/\hat{g}}\,.
\end{align}
These equations match those of ref.~\cite{Arutyunov:2012ai} even if their construction uses the inverse of our S~matrix, which can be seen from the eigenvalues of the transfer matrix.

A given eigenstate of the S matrix is characterised by the integers $K^\mathrm{I}, K^\mathrm{II}_\pm$ and $K^\mathrm{III}_\pm$, which, as is transparent from the construction of the Bethe ansatz, are counting the number of excitations. 
It is convenient to relate them to the charges of the eigenstate under the Cartans of $\su(4)$, which corresponds to the Dynkin labels $[q_1,p,q_2]$, and under the Cartans of $\su(2,2)$, given by $[s_+,r,s_-]$.
Our convention will follow the standard one for the Bethe ansatz in the $\su(2)$ grading (in the notation of~\cite{Beisert:2005fw}, this is $\eta_+=\eta_-=1$), and we indicate it here for the sake of completeness.
Denoting by $L^\alpha{}_\beta$, $L^{\dot{\alpha}}{}_{\dot{\beta}}$, $R^a{}_b$ and $R^{\dot{a}}{}_{\dot{b}}$ the eigenvalues under the Cartan generators $\mathbf{L}^\alpha{}_\beta$,  $\mathbf{L}^{\dot{\alpha}}{}_{\dot{\beta}}$, $\mathbf{R}^a{}_b$ and $\mathbf{R}^{\dot{a}}{}_{\dot{b}}$, we define
\begin{equation} \begin{aligned}
    q_+ &= R^{4}{}_4 - R^{3}{}_3\,, \qquad &q_- &= R^{\dot{4}}{}_{\dot{4}} - R^{\dot{3}}{}_{\dot{3}}\,, \qquad &p &= R^{\dot{3}}{}_{\dot{3}} - R^{4}{}_4 \,, \\
    s_+ &=  L^2{}_2 - L^1{}_1\,, &\qquad s_- &= L^{\dot{2}}{}_{\dot{2}} - L^{\dot{1}}{}_{\dot{1}}\,,
    \qquad
    &r&= -D+ L^1{}_1+L^{\dot{1}}{}_{\dot{1}}\,,
    \end{aligned}
\end{equation}
where $D$ is the dilatation operator. The vacuum has charges $D=p=J$ while the other label are zero, so that
\begin{equation}
    q_+ = K^\mathrm{I}-K_+^\mathrm{II}\,, \qquad
    p = J - 2 K^\mathrm{I} + K_+^\mathrm{II}+ K_-^\mathrm{II}\,,\qquad
    q_- = K^\mathrm{I}-K_-^\mathrm{II}\,,
\end{equation}
and
\begin{equation}
    s_+ = K^\mathrm{II}_+ - 2 K^\mathrm{III}_+\,,\qquad   
    s_- = K^\mathrm{II}_--2 K^\mathrm{III}_-\,,
\end{equation}
while $r$ is given in terms of the non-quantised charge $D$ which in turn is fixed in terms of the lightcone energy,
\begin{equation}
    D-\frac{q_++2p+q_-}{2}= 
    H
    =\sum_{j=1}^{K^{I}} H(p_j)\,,
\end{equation}
where the dispersion relation follows from the closure condition~\eqref{eq:closure}
\begin{equation}
     \sinh^2\left(\frac{a H(p) }{2}\right) = - \xi^2 \sin^2\left(\frac{p}{2}\right)+(1-\xi^2) \sinh^2\left(\frac{a}{2}\right)\,, \qquad q=e^{-a}\,.
\end{equation}

\section*{Acknowledgments}
We thank Sergey Frolov, Ben Hoare, Arkady Tseytlin and Stijn van Tongeren for useful discussions and comments. The work of FS is supported by the Swiss National Science Foundation via the Early Postdoc.Mobility fellowship ``$q$-deforming AdS/CFT''.

\appendix
\section{Coordinate Bethe ansatz for the Fermionic S matrix}
\label{app:CBA}
In this appendix we work out the coordinate Bethe ansatz for the (rescaled) fermionic S matrix $\mathcal R'$ defined in section~\ref{sec:Smatrix:Fermionicrescaled}.  We will restrict to states of length two, generalisation to states of higher length follows from factorisation of the S matrix.
\subsection{Vacuum}
In view of the form of the two-body S matrix a good choice for the level-II vacuum is the homogeneous state
\begin{equation}
\ket{0}^{\mathrm{II}} = \ket{\phi_3(x_1) \phi_3(x_2)}~.
\end{equation}
The vacuum is an eigenstate of the S matrix,
\begin{equation}
\check{\mathcal{R}}' \ket{0}^\mathrm{II}  = \check{\mathcal{R}}' \ket{\phi_3(x_1) \phi_3(x_2)} = A \ket{\phi_3(x_2) \phi_3(x_1)} = R^\mathrm{I,I} \ket{0}^\mathrm{II}_\pi~,
\end{equation}
with eigenvalue $R^\mathrm{I,I} = A$. Here $\pi={(1,2)}$ denotes the permutation of rapidities $x_1$ and $x_2$.
\subsection{Propagation}
For a state with one excitation above the level two vacuum we make the ansatz ($\alpha=1,2$)
\begin{align}
\ket{\psi_\alpha(y)}^{\mathrm{II}} &= f(y,x_1) \ket{\psi_\alpha(x_1) \phi_3(x_2)} + f(y,x_2) R^{\mathrm{II,I}}(y,x_1) \ket{\phi_3(x_1) \psi_\alpha(x_2)} ~, \\
\ket{\psi_\alpha(y)}^{\mathrm{II}}_{\pi} &= f(y,x_2) \ket{\psi_\alpha(x_2) \phi_3(x_1)} + f(y,x_1) R^{\mathrm{II,I}}(y,x_2) \ket{\phi_3(x_2) \psi_\alpha(x_1)} ~.
\end{align}
The compatibility condition
\begin{equation}
\check{\mathcal{R}}' \ket{\psi_\alpha(y)}^{\mathrm{II}} = A \ket{\psi_\alpha(y)}^{\mathrm{II}}_\pi~,
\end{equation}
then gives rise to the equations
\begin{align}
f(y,x_1) K + f(y,x_2) R^{\mathrm{II,I}}(y,x_1) G &= A f(y,x_2)~, \\
f(y,x_1) L + f(y,x_2 )R^{\mathrm{II,I}}(y,x_1) H &= A f(y,x_1) R^{\mathrm{II,I}}(y, x_2)~.
\end{align}
These are solved by 
\begin{equation}
f(y,x_j) =  \frac{y \gamma_j}{y-x_j^+} ~,  
\end{equation}
and
\begin{equation}
R^{\mathrm{II,I}}(y, x_j)= q^{1/2} U_j V_j \frac{y-x_j^-}{y-x_j^+}~.
\end{equation}
A one-particle change of basis of the S matrix (the rescaling we did) affects the function $f(y,x_j)$ (and in fact we might have different functions for each type of excitation, in particular here we could have two different functions $f_1(y,x_j)$ and $f_2(y,x_j$), the change of basis we did precisely avoids that, and we are left with exactly the same propagation equations as in the distinguished case), but the function $R^\mathrm{II,I}$, which is the important one in the construction of the Bethe equations, remains the same.

\subsection{Scattering}
For a state with two excitations above the vaccum we make the ansatz
\begin{align}
\ket{\psi_\alpha(y_1) \psi_\beta(y_2)}^{\mathrm{II}} &= f(y_1,x_1) f(y_2,x_2) R^{\mathrm{II,I}}(y_2, x_1) \ket{\psi_\alpha(x_1) \psi_\beta(x_2)}~, \\
\ket{\psi_\alpha(y_1) \psi_\beta(y_2)}^{\mathrm{II}}_\pi &= f(y_1,x_2) f(y_2,x_1) R^{\mathrm{II,I}}(y_2, x_2) \ket{\psi_\alpha(x_2) \psi_\beta(x_1)}~.
\end{align}
By construction the excitation with rapidity $y_1$ is always to the left of the one with rapidity $y_2$. We additionally have the freedom to exchange them. For this purpose we introduce the second level S matrix $\mathcal{R}^{\mathrm{II}}$, as well as $\check{\mathcal{R}}^{\mathrm{II}}= \Pi^g \mathcal{R}^{\mathrm{II}}$. Motivated by the results from the algebraic Bethe ansatz we assume that this level II S matrix has $\su(2)_q$ symmetry but allow for a rapidity-dependent one-particle change of basis,
\begin{equation}
\begin{aligned}
\label{eq:levelIIR_f}
\mathcal{R}^{\mathrm{II}} \ket{\psi_1(y_1) \psi_1(y_2)}^{\mathrm{II}} &= M \ket{\psi_1(y_1) \psi_1(y_2)}^{\mathrm{II}}~, \\
\mathcal{R}^{\mathrm{II}} \ket{\psi_1(y_1) \psi_2(y_2)}^{\mathrm{II}} &= \frac{M-N}{q+q^{-1}} \ket{\psi_1(y_1)\psi_2(y_2) }^{\mathrm{II}}+\frac{h(y_1)}{h(y_2)}\frac{q M + q^{-1} N}{q+q^{-1}}\ket{ \psi_2(y_1) \psi_1(y_2)}^{\mathrm{II}}~, \\
\mathcal{R}^{\mathrm{II}} \ket{\psi_2(y_1) \psi_1(y_2)}^{\mathrm{II}} &=  \frac{h(y_2)}{h(y_1)}\frac{q^{-1} M + q N}{q+q^{-1}}\ket{ \psi_1(y_1) \psi_2(y_2)}^{\mathrm{II}}+\frac{M-N}{q+q^{-1}} \ket{\psi_2(y_1) \psi_1(y_2) }^{\mathrm{II}}~, \\
\mathcal{R}^{\mathrm{II}} \ket{\psi_2(y_1) \psi_2(y_2)}^{\mathrm{II}} &= M \ket{\psi_2(y_1) \psi_2(y_2) }^{\mathrm{II}}~,
\end{aligned}
\end{equation}
with the shorthand notation $M=M(y_1,y_2)$ and $N=N(y_1,y_2)$. The notation is motivated by the fact that, as we will see, these functions are precisely the same as their homonyms derived in the context of the algebraic Bethe ansatz. 
The ansatz for the two-excitation state then reads
\begin{equation}
\ket{\psi_\alpha \psi_\beta}^\mathrm{II}  = \ket{\psi_\alpha(y_1) \psi_\beta(y_2)}^{\mathrm{II}} + \check{\mathcal{R}}^{\mathrm{II}} \ket{\psi_\alpha(y_1) \psi_\beta(y_2)}^{\mathrm{II}}~. 
\end{equation}
At this point we must also include the $\ket{\phi_4}$ states, which behave as a double excitation above the $\ket{\phi_3}$ vacuum. For this we introduce
\begin{equation}
\begin{aligned}
\ket{\phi_{4,\alpha \beta}}^\mathrm{II} = & f(y_1,x_1) f(y_2, x_1) f_{\alpha \beta}(y_1, y_2, x_1) \ket{\phi_4(x_1) \phi_3(x_2)} \\
&+ f(y_1, x_2)  R^{\mathrm{II,I}}(y_1,x_1)  f(y_2, x_2)R^{\mathrm{II,I}}(y_2,x_1) f_{\alpha \beta}(y_1, y_2, x_2) \ket{\phi_3(x_1) \phi_4(x_2)}~.
\end{aligned}
\end{equation}
and
\begin{align}
\begin{aligned}
\ket{\phi_{4,\alpha \beta}}^\mathrm{II}_\pi = & f(y_1,x_2) f(y_2, x_2) f(y_1, y_2, x_2) \ket{\phi_4(x_2) \phi_3(x_1)} \\
&+ f(y_1, x_1)  R^{\mathrm{II,I}}(y_1,x_2)  f(y_2, x_1)R^{\mathrm{II,I}}(y_2,x_2) f(y_1, y_2, x_1) \ket{\phi_3(x_2) \phi_4(x_1)}~.
\end{aligned}
\end{align}
In contrast to the distinguished case we allow for two different fusion functions $f_{12}(y_1,y_2,x)$ and $f_{21}(y_1,y_2,x)$. 
The full ansatz for the state with two excitations then reads  (no sum over $\alpha$ and $\beta$)
\begin{equation}
\ket{\psi_{\alpha \beta}}^{\mathrm{II}} = \ket{\psi_\alpha \psi_\beta}^\mathrm{II} + C_{\alpha \beta} \ket{\phi_{2,\alpha \beta}}^\mathrm{II}~,
\end{equation}
with $C_{33}=C_{44}=0$, and without loss of generality we fix $C_{34}=C_{43}=1$ (a constant can always be reabsorbed into the unknown functions $f_{\alpha \beta}(y_1,y_2,x)$). 

The compatibility condition is then solved for the functions $M$, $N$ and $h$ as in the algebraic Bethe ansatz, namely
\begin{align}
    M(y_1,y_2) &=1~, \\
    N(y_1,y_2) &= \frac{q^{-1} u_1-q u_2-i/\hat{g}}{q^{-1} u_2 - q u_1 - i/\hat{g}}~, \qquad u=y+\frac{1}{y}~, \\
    h(y) &= \frac{y}{y+\xi}~,
\end{align}
while the auxiliary functions take the form
\begin{align}
    f_{12}(y_1,y_2,x) &=  q^{-1} (y_1 - 
   y_2)  \frac{x_1^+}{\gamma_1^2} \frac{q (x_1^+ - x_1^-) \xi + (q - q^{-1}) y_1 \xi (1 + 
      x_1^+ \xi) + 
   \tilde{y}}{q y_2 (y_1 + \xi) (1 + 
      y_1 \xi) - q^{-1} y_1 (y_2 + \xi) (1 + y_2 \xi))}\,, \\
    f_{21}(y_1,y_2,x) &= -  (y_1 - 
   y_2)  \frac{x_1^+}{\gamma_1^2}  \frac{ q^{-1} (x_1^+ - x_1^-) \xi + (q - q^{-1}) y_2 \xi (1 + 
      x_1^- \xi) + 
   \tilde{y}}{ q y_2 (y_1 + \xi) (1 + 
      y_1 \xi) - q^{-1} y_1 (y_2 + \xi) (1 + y_2 \xi))}\,, \\
      \tilde{y} &= y_1 y_2  \big(q (1 + x_1^- \xi) - 
      q^{-1} (1 + x_1^+ \xi)\big)\,.
\end{align}

\subsection{Final level}
We are now left with diagonalising the level-II S matrix $\check{\mathcal{R}}^\mathrm{II}$.
We choose as level-III vacuum $\ket{0}^\mathrm{III}=\ket{\psi_1(y_1) \psi_1(y_2)}^\mathrm{II}$, with $\check{\mathcal{R}}^\mathrm{II} \ket{0}^\mathrm{III} = -M \ket{0}^\mathrm{III}_\pi \rightarrow R^\mathrm{II,II}=-M$. Very similarly to what we had in the propagation case above, the state with one excitation above this vacuum reads
\begin{align}
\label{eq:levelII-ordering}
\ket{\psi_2(v)}^{\mathrm{III}} &= f'(v,y_1) \ket{\psi_2(y_1) \psi_1(y_2)}^\mathrm{II} + f'(v,y_2) R^{\mathrm{III,II}}(v,y_1) \ket{\psi_1(y_1) \psi_2(y_2)}^\mathrm{II} ~, \\
\ket{\psi_2(v)}^{\mathrm{III}}_{\pi} &= f'(v,y_2) \ket{\psi_2(y_1) \psi_1(y_2)}^\mathrm{II} + f'(v,y_1) R^{\mathrm{III,II}}(v,y_2) \ket{\psi_1(y_1) \psi_2(y_2)}^ \mathrm{II} ~.
\end{align}
The compatibility condition now reads
\begin{equation}
\check{\mathcal{R}}^\mathrm{II} \ket{\psi_2(v)}^\mathrm{III} = -M \ket{\psi_2(v)}^\mathrm{III}_\pi~,
\end{equation}
and yields the two equations
\begin{align}
\frac{h(y_2)}{h(y_1)}\frac{q^{-1} M + q N }{q+q^{-1}} f'(v,y_1) + \frac{M-N}{q+q^{-1}} f'(v, y_2) R^\mathrm{III,II}(v,y_1) &= -M f'(v,y_2)~, \\
 \frac{M-N}{q+q^{-1}} f'(v,y_1) + \frac{h(y_1)}{h(y_2)}\frac{q M + q^{-1} N }{q+q^{-1}} f'(v, y_2) R^\mathrm{III,II}(v,y_1) &= -M  f'(v,y_1) R^\mathrm{III,II}(v,y_2)~.
\end{align}
As we can see, the one-particle change of basis in the auxiliary S matrix can be reabsorbed into the definition of the auxiliary function, which becomes
\begin{equation}
f'(v,y) = - h(y) \frac{q^2 (y+\xi)(v+2 \xi)(1+y \xi)}{(\xi-\xi^{-1})(\xi+y^2 \xi + y(1+\xi^2-q^2(1+v \xi + \xi^2)))}~,
\end{equation}
while the important piece for the Bethe equations remains invariant,
\begin{equation}
R^\mathrm{III,II} (v,u)= \frac{u-v}{q^{-1} u - q v - i/\hat{g}}~, \qquad u=y+\frac{1}{y}~.
\end{equation}
Finally, the ansatz for the state with two $v$-type excitations reads
\begin{equation}
\ket{\psi_2 \psi_2}^\mathrm{III} = \ket{\psi_2(v_1) \psi_2(v_2)}^\mathrm{III} + \check{\mathcal{R}}^\mathrm{III} \ket{\psi_2(v_1) \psi_2(v_2)}^\mathrm{III} ~, 
\end{equation}
where  
\begin{equation}
\ket{\psi_2(v_1) \psi_2(v_2)}^\mathrm{III} = f'(v_1,y_1) f'(v_2,y_2) R^{\mathrm{III,II}}(v_2, y_1) \ket{\psi_2(y_1) \psi_2(y_2)}^\mathrm{II}~,
\end{equation}
and the level-III S matrix $\check{\mathcal{R}}^\mathrm{III}$ governs the scattering of two $v$-type excitations, and simply takes the form
\begin{equation}
\check{\mathcal{R}}^\mathrm{III} \ket{\psi_2 (v_1)\psi_2(\omega_2)}^\mathrm{III} = R^\mathrm{III,III}\ket{\psi_2 (v_2)\psi_2(v_1)}^\mathrm{III} ~.
\end{equation}
The compatibility condition
\begin{equation}
\check{\mathcal{R}}^\mathrm{II} \ket{\psi_2 \psi_2}^\mathrm{III} = -M \ket{\psi_2 \psi_2}^\mathrm{III}_\pi~,
\end{equation}
gives rise to a single equation
\begin{equation}
\begin{aligned}
&f'(v_1, y_1) f'(v_2, y_2) R^\mathrm{III,II}(v_2, y_1) + R^\mathrm{III,III}(v_1, v_2) f'(v_2, y_1) f'(v_1, y_2) R^\mathrm{III,II}(v_1, y_1) \\
&=f'(v_1, y_2) f'(v_2, y_1) R^\mathrm{III,II}(v_2, y_2) + R^\mathrm{III,III}(v_1, v_2) f'(v_2, y_2) f'(v_1, y_1) R^\mathrm{III,II}(v_1, y_2) ~.
\end{aligned}
\end{equation}
The rescaling of the auxiliary function $f'$ drops out and the solution reads as in the distinguished case,
\begin{equation}
R^\mathrm{III,III}(v_1,v_2)= \frac{q^{-1} v_2 - q v_1 - i/\hat{g}}{q v_2 - q^{-1} v_1 + i/\hat{g}}~.
\end{equation}
\subsection{Bethe equations}
In the context of the coordinate Bethe ansatz, the Bethe equations arise from imposing periodicity of the wavefunction. For instance, in ansatz \eqref{eq:levelII-ordering} we chose an ordering of the level-$\mathrm{II}$ excitations, requiring that $y_1$ is always to the left of $y_2$ and introduced a scattering matrix to pass one excitation through the other. But using periodicity we could also have started with an ansatz where $y_2$ is always to the left of $y_1$. Different choices can also be made for the level $\mathrm{III}$ excitations. Requiring that the two procedures yield the same result imposes, for a state with $K^\mathrm{I}$ level-$\mathrm{I}$ excitations, $K^\mathrm{II}$ level-$\mathrm{II}$ excitations and $K^\mathrm{III}$ level-$\mathrm{III}$ excitations, the equations
\begin{align}
1 &= \prod_{j=1}^{K^\mathrm{I}} R^\mathrm{I,II}(x_j,u_k) \prod_{j=1}^{K^\mathrm{III}} R^\mathrm{III,II}(v_j,u_k)\,, \qquad k=1,\dots,K^\mathrm{II}\,, \\
1 &=\prod_{j=1}^{K^\mathrm{II}} R^\mathrm{II,III}(u_j, v_k) \prod_{l \neq k}^{K^\mathrm{III}} R^\mathrm{III,III}(v_l,v_k)\,, \qquad k=1,\dots,K^\mathrm{III}\,, 
\end{align}
where we recall that
\begin{equation}
R^\mathrm{III,III}(v_j,v_k) =  \frac{q^{-1} v_k-q v_j-i/\hat{g}}{q v_k -q^{-1} v_j+i/\hat{g}}~, \qquad R^\mathrm{II,III}(u_j,v_k) = \frac{q^{-1}u_j-q v_k - i/\hat{g}}{u_j-v_k}\,, 
\end{equation}
\begin{equation}
 \qquad R^\mathrm{II,I}(u_k,x_j)= q^{1/2} U_j V_j \frac{y_k - x_j^-}{y_k-x_j^+}\,.
\end{equation}
These are precisely the same equations as the ones arising from the algebraic Bethe ansatz.

\newpage
\bibliographystyle{JHEP}
\bibliography{refs}

\end{document}